\documentclass[aps,prd,twocolumn,superscriptaddress,floatfix,showpacs,showkeys]{revtex4-1}

\usepackage[utf8]{inputenc}
\usepackage[T1]{fontenc}
\usepackage{slashed}
\usepackage{times}

\usepackage{amssymb,amsfonts,amsmath,amsthm}
\usepackage{dsfont,bbm}
\usepackage{dcolumn}

\usepackage{graphicx}
\usepackage[caption=false]{subfig}
\usepackage{color}

\newcommand{\MS}{{\overline{\mathrm{MS}}}}
\newcommand{\cO}{\mathcal {O}}
\newcommand{\Dl}[1]{\overset{\leftarrow}{D}_{#1}}
\newcommand{\Dr}[1]{\overset{\rightarrow}{D}_{#1}}
\newcommand{\Dd}[1]{\overset{\leftrightarrow}{D}_{#1}}

\begin{document}

\title{Second Moment of the Pion Light-cone Distribution Amplitude from Lattice QCD}

\author{V.M.~Braun}
   \affiliation{Institut f\"ur Theoretische Physik, Universit\"at Regensburg, 93040 Regensburg, Germany}
\author{S.~Collins}
   \affiliation{Institut f\"ur Theoretische Physik, Universit\"at Regensburg, 93040 Regensburg, Germany}

\author{M.~{G\"ockeler}}
   \affiliation{Institut f\"ur Theoretische Physik, Universit\"at Regensburg, 93040 Regensburg, Germany}

\author{P.~{P\'erez-Rubio}}
   \email{Paula.Perez-Rubio@ur.de}
   \affiliation{Institut f\"ur Theoretische Physik, Universit\"at Regensburg, 93040 Regensburg, Germany}

\author{A.~{Sch\"afer}}
    \affiliation{Institut f\"ur Theoretische Physik, Universit\"at Regensburg, 93040 Regensburg, Germany}
\author{R.~W.~{Schiel}}
   \affiliation{Institut f\"ur Theoretische Physik, Universit\"at Regensburg, 93040 Regensburg, Germany}
\author{A.~{Sternbeck}}
    \affiliation{Theoretisch-Physikalisches Institut,
     Friedrich-Schiller-Universit\"at Jena, 07743 Jena, Germany}
    \affiliation{Institut f\"ur Theoretische Physik, 
     Universit\"at Regensburg, 93040 Regensburg, Germany}


\begin{abstract}
We present the results of a lattice study of the second moment of the 
light-cone pion distribution amplitude using two flavors of 
dynamical (clover) fermions on lattices of different volumes and 
pion masses down to $m_\pi\sim 150 \, \mathrm {MeV}$. At lattice spacings
between $0.06 \, \mathrm {fm}$ and $0.08 \, \mathrm {fm}$ we find for
the second Gegenbauer moment the value $a_2 = 0.1364(154)(145)$ at the 
scale $\mu=2 \, \mathrm {GeV}$ in the $\MS$ scheme, where the first error 
is statistical including the uncertainty of the chiral extrapolation, 
and the second error is the estimated uncertainty coming from the 
nonperturbatively determined renormalization factors.
\end{abstract}


\maketitle

\section{Introduction}

Hard exclusive processes involving energetic pions in the final state are 
sensitive to the momentum fraction distribution of the valence quarks 
at small transverse separations, usually called the pion distribution 
amplitude (DA). Classical 
applications~\cite{Chernyak:1977as,Radyushkin:1977gp,Lepage:1979zb}
have been to exclusive two-photon processes, e.g., the pion 
electromagnetic form factor at large momentum transfer and the 
transition form factor $\gamma^*\to \pi\gamma$. The latter process 
plays a very special r\^ole as the simplest hard exclusive reaction 
where QCD factorization can be tested at a quantitative level. It 
received a lot of interest recently, triggered by the partially 
conflicting measurements by BaBar~\cite{BABAR} and BELLE~\cite{Uehara:2012ag}
up to photon virtualities of the order of $40 \, \mathrm {GeV}^2$, 
see, e.g.,~\cite{Radyushkin:2009zg,Agaev:2010aq,Bakulev:2012nh,Agaev:2012tm,Chernyak:2014wra}.
Arguably, the most important application of the pion DA is currently the study 
of semileptonic weak decays $B\to\pi\ell\bar\nu_\ell$  at large 
recoil~\cite{Ball:2004ye,Duplancic:2008ix,Khodjamirian:2011ub}  
using light-cone sum rules (LCSR)~\cite{Balitsky:1989ry,Chernyak:1990ag} 
and weak hadronic decays $B\to \pi\pi$ etc.\ in the framework of QCD 
factorization~\cite{Beneke:1999br,Beneke:2003zv}. 
Both reactions contribute prominently to the determination of parameters of 
the quark mixing matrix in the Standard Model. 
     
The precise definition of the pion DA $\phi_\pi(x,\mu^2)$ is based on the 
representation~\cite{Chernyak:1977as,Radyushkin:1977gp,Lepage:1979zb}
as the matrix element of a nonlocal light-ray quark-antiquark operator.
For example, for a positively charged pion
\begin{eqnarray}
\lefteqn{ \langle 0| \bar d(z_2n)\slashed{n}\gamma_5 [z_2n,z_1n] u(z_1n) |\pi(p)\rangle }
\nonumber\\ &=& if_\pi (p\cdot n) 
 \int_0^1 dx\, e^{-i(z_1 x + z_2 (1-x)) p\cdot n}\phi_{\pi}(x,\mu^2)\,, 
\label{def:phi}
\end{eqnarray}
where $p^\mu$ is the pion momentum, $n^\mu$ is a light-like vector, 
$n^2=0$, $z_{1,2}$ are real numbers, $[z_2n,z_1n]$ is the Wilson line 
connecting the quark and the antiquark fields and 
$f_\pi=132 \, \mathrm {MeV}$ is the 
usual pion decay constant. The DA $\phi_{\pi}(x,\mu^2)$ is scale-dependent, 
which is indicated by the argument $\mu^2$.

The physical interpretation of the variable $x$ is that the $u$-quark
carries the fraction $x$ of the pion momentum, so that $1-x$ is the
momentum fraction carried by the $\bar d$-antiquark.
Neglecting isospin breaking effects and electromagnetic corrections
the pion DA is symmetric under the interchange $x \leftrightarrow 1-x$:
\begin{equation}
   \phi_\pi(x,\mu^2) = \phi_\pi(1-x,\mu^2) \,.
\end{equation} 
Due to this symmetry, only the even moments involving the momentum fraction 
\emph{difference} 
\begin{align}
   \xi = x - (1-x) = 2x-1
\end{align}
carry nontrivial physical information:
\begin{align}
 \langle \xi^n \rangle = \int_0^1 dx\, (2x-1)^n \phi_\pi(x,\mu^2)\,, && n=0,2,\ldots.
\end{align}
The definition in (\ref{def:phi}) implies the normalization condition
\begin{equation}
   \int_0^1 dx\, \phi_{\pi}(x,\mu^2) = 1\,.
\end{equation}

A convenient parameterization of DAs is provided by the conformal 
expansion~\cite{Brodsky:1980ny,Makeenko:1980bh,Braun:2003rp}.
The underlying idea is to use the conformal symmetry of the QCD Lagrangian 
to separate transverse and longitudinal variables in the 
light-front pion wave function, similar in spirit to the 
partial-wave decomposition in quantum mechanics.
The dependence on transverse coordinates is formulated as a scale
dependence of the relevant operators and is governed by
renormalization-group equations. 
The dependence on the longitudinal momentum fractions is described in
terms of Gegenbauer polynomials $C^{3/2}_n(2x-1)$ which correspond to
irreducible representations of the collinear conformal group SL(2,$\mathbb R$).
In this way one obtains
\begin{equation}
\phi_\pi(x,\mu^2) = 6 x(1-x) \left[1 + \sum\limits_{n=2,4,\ldots}^\infty
  a_{n}(\mu^2) C_{n}^{3/2}(2x-1)\right],
\label{eq:gegen1}
\end{equation}
where all nonperturbative information is contained in the set of 
coefficients (Gegenbauer moments)  $a_n(\mu_0^2)$ 
at a certain reference scale $\mu_0$. To leading-logarithmic accuracy (LO), 
the Gegenbauer moments renormalize multiplicatively with the anomalous 
dimensions rising slowly with $n$. Thus the higher-order contributions 
in the Gegenbauer expansion are suppressed at large
scales and asymptotically only the leading term survives, 
\begin{equation}
  \phi^{\mathrm {as}}_\pi(x) = 6 x(1-x)\,,
\label{asymptotic}
\end{equation}
which is usually referred to as the asymptotic pion DA.
It is widely accepted, however, that the pion DA deviates significantly
from its asymptotic form at scales that can be achieved in 
experiments.

A particular model of the pion DA proposed by Chernyak and
Zhitnitsky in 1982~\cite{Chernyak:1981zz} has played an important 
r\^ole in historic perspective. It was based on a calculation of $a_2$ 
using QCD sum rules~\cite{Shifman:1978bx}, which
resulted in a large value $a_2 \sim 0.5-0.6$ (at the scale 1 GeV),  
and the assumption that all higher-order coefficients can be neglected. 

Since then, different approaches have been used: 
QCD sum rules with various improvements 
(e.g.~\cite{Bakulev:2001pa,Khodjamirian:2004ga,Ball:2006wn}), 
LCSR-based analysis of experimental data on the pion electromagnetic 
and transition form factors 
(e.g.~\cite{Agaev:2010aq,Bakulev:2012nh,Agaev:2012tm}) and weak 
$B$-meson decay form factors (e.g.~\cite{Khodjamirian:2011ub}), 
lattice calculations~\cite{Braun:2006dg,Arthur:2010xf}
and recently also in the framework of Dyson-Schwinger 
equations~\cite{Chang:2013pq}. A recent compilation of the 
existing results 
for $a_2$ can be found in Table~1 of Ref.~\cite{Agaev:2010aq}. 

Estimates of yet higher-order Gegenbauer coefficients are rather uncertain. 
A direct calculation of $a_4$ proves to be difficult and its extraction 
from the experimental data on, e.g., the pion transition form factor 
is complicated by the fact the LO contribution is proportional to the 
sum of Gegenbauer moments 
\begin{equation}
 \int_0^1 \frac{dx}{x} \phi_\pi(x,\mu^2) 
                 = 3[1 + a_2(\mu^2)+ a_4(\mu^2)+\ldots]\,.
\end{equation} 
Thus, the values of $a_2(\mu^2)$ and $a_4(\mu^2)$ obtained in these 
extractions appear to be strongly correlated. The strong scaling 
violation in the pion transition form factor observed by BaBar~\cite{BABAR}
(but not confirmed by BELLE~\cite{Uehara:2012ag}) would imply a 
considerable enhancement of the pion DA close to the end-points, 
meaning that the expansion in Gegenbauer polynomials is converging
very slowly if at all, see the detailed discussion 
in~\cite{Radyushkin:2009zg,Agaev:2010aq,Agaev:2012tm}.  
The forthcoming upgrade of the Belle experiment and the 
KEKB accelerator~\cite{Adachi:2014kka}, which aims to increase 
the experimental data set by a factor of 50, will allow one to 
measure transition form factors and related observables with unprecedented 
precision and resolve this issue. The question at stake is whether 
hard exclusive hadronic reactions are under theoretical control, 
which is highly relevant for all future high-intensity, medium energy 
experiments like, e.g., PANDA. On the theory side, several proposals 
exist how it might be possible to access DA moments beyond the second 
one (or the DA pointwise in $x$) on the lattice, 
e.g.,~\cite{Braun:2007wv,Ji:2013dva}, but the corresponding techniques 
are only in the exploratory stage.

In this work we extend the lattice study~\cite{Braun:2006dg} of the 
second moment of the pion DA  by making use of a larger set of lattices 
with different volumes, lattice spacings and pion masses down to 
$m_\pi\sim 150 \, \mathrm {MeV}$ and implementing several 
technical improvements.
We employ the variational approach with two and three interpolators 
to improve the signal from the pion state. 
The renormalization of the lattice data is performed nonperturbatively
utilizing a version of the RI'-SMOM scheme. For the first time we include
a nonperturbative calculation of the renormalization factor corresponding
to the mixing with total derivatives, which proves to have a significant 
effect. Our main result is 
\begin{equation}
 a_2 = 0.1364(154)(145)(?)
\label{result-a2}
\end{equation}
for the second Gegenbauer moment of the pion DA, and 
\begin{equation}
\langle \xi^2 \rangle = 0.2361(41)(39)(?) \,.
\label{result-xi2}
\end{equation}
Both numbers refer to the scale $\mu = 2 \, \mathrm {GeV}$ in the $\MS$ scheme. 
The first error combines the statistical uncertainty and the 
uncertainty of the chiral extrapolation.
The second error is the estimated uncertainty contributed by the
nonperturbative determination of the renormalization and mixing
factors. Our lattice data are collected for the lattice spacing 
$a = 0.06-0.08 \, \mathrm {fm}$, and this range is not large enough to 
ensure a  reliable continuum extrapolation. 
The corresponding remaining uncertainty is indicated as (?). 
It has to be addressed in a future study.

The paper is organized as follows. In the next section we
discuss the aspects of the continuum description of the pion DA that
are relevant for our work. The basics of the lattice formulation are 
given in Sec.~\ref{sect:latform}. An important ingredient in our calculation
is the nonperturbative evaluation of the renormalization and mixing 
coefficients, which is described in Sec.~\ref{sect:renco}. The methods
applied in the analysis of the bare data are detailed in 
Sec.~\ref{sect:anadata}. Our results are presented in 
Sec.~\ref{sect:results}, followed by our conclusions and an outlook. 
In an Appendix we collect Tables of intermediate results for each gauge 
field ensemble used in our work.

\section{Moments of the pion distribution amplitude}

The nonlocal operator in the expression for the pion DA (\ref{def:phi}) 
is defined as a generating function for renormalized leading-twist 
(i.e., twist two) local operators,
\begin{eqnarray}
\lefteqn{\hspace*{-1cm}\bar d(z_2n)\slashed{n}\gamma_5 [z_2n,z_1n] u(z_1n) =}
\nonumber\\ 
&=& 
\sum_{k,l=0}^\infty \frac{z_2^k z_1^l}{k!l!} 
n^\rho n^{\mu_1}\ldots n^{\mu_{k+l}} 
\mathcal{M}_{\rho\mu_1\ldots\mu_{k+l}}^{(k,l)} \,,
\end{eqnarray}
where 
\begin{eqnarray}
\lefteqn{ \mathcal{M}_{\rho\mu_1\ldots\mu_{k+l}}^{(k,l)} =} 
\nonumber\\
&=& \bar d(0)\Dl{(\mu_1}\ldots \Dl{\mu_k}
\Dr{\mu_{k+1}}\ldots \Dr{\mu_{k+l}}\gamma_{\rho)}\gamma_5 u(0)\,. 
\end{eqnarray}
Here $D_\mu$ is the covariant derivative and $(\ldots)$ denotes the 
symmetrization of all enclosed Lorentz indices and the subtraction of 
traces. The local operators $\mathcal{M}_{\rho\mu_1\ldots\mu_{k+l}}^{(k,l)}$ 
are assumed to be renormalized, e.g., in the $\MS$\ scheme.   

As a consequence, moments of the pion DA are given by matrix elements 
of local operators:
\begin{equation}
i^{k+l} \langle 0 |\mathcal{M}_{\rho\mu_1\ldots\mu_{k+l}}^{(k,l)} 
 | \pi (p) \rangle =
 i f_\pi p_{ ( \rho}p_{\mu_1}\ldots p_{\mu_{k+l})} 
    \langle x^l(1\!-\!x)^k \rangle.
\end{equation}
Neglecting isospin breaking effects and electromagnetic corrections 
one obtains the symmetry relation 
\begin{equation}
 \langle 0 |\mathcal{M}_{\rho\mu_1\ldots\mu_{k+l}}^{(k,l)} | \pi (p) \rangle = 
\langle 0 |\mathcal{M}_{\rho\mu_1\ldots\mu_{k+l}}^{(l,k)} | \pi (p) \rangle 
\end{equation}
and thus
\begin{equation}
 \langle x^l(1\!-\!x)^k\rangle  = \langle x^k(1\!-\!x)^l\rangle. 
\end{equation}
In addition, the product (Leibniz) rule for derivatives 
\begin{equation}
  \mathcal{M}_{\rho\mu_1\ldots\mu_{k+l+1}}^{(k+1,l)} 
 + \mathcal{M}_{\rho\mu_1\ldots\mu_{k+l+1}}^{(k,l+1)} 
= \partial_{(\mu_{k+l+1}} \mathcal{M}_{\rho\mu_1\ldots\mu_{k+l})}^{(k,l)}   
\label{leibniz}
\end{equation}
gives rise to the momentum-conservation constraint
\begin{equation}
 \langle x^{l+1}(1\!-\!x)^k\rangle +  \langle x^{l}(1\!-\!x)^{k+1}\rangle 
 = \langle x^{l}(1\!-\!x)^{k}\rangle\,.   
\end{equation}
Specializing to the second moment, $l+k=2$, it is easy to see that 
only one independent matrix element remains, e.g.,
\begin{equation}
 \langle \xi^2 \rangle  =  1 - 4 \langle x(1-x)\rangle
\label{relations1a}
\end{equation}
or
\begin{align}
 a_2 & = \frac{7}{18}\langle C^{3/2}_2(2x-1)\rangle
    =   \frac{7}{12}\big[5\langle \xi^2\rangle -1\big] 
\notag\\
&= \frac{7}{3}\big[1 - 5 \langle x(1-x)\rangle \big]\,,
\label{relations1b}
\end{align}
so that any moment $\langle \xi^2\rangle$, $a_2$, $\langle x(1-x)\rangle$ 
etc.\ can be used as a nonperturbative parameter to characterize the 
shape of the pion DA. Lacking any \emph{a priori}\ information on the 
relative size of the different contributions, all such choices
are equivalent. It is widely expected, however, that the numerical 
value of $\langle \xi^2\rangle$ is not far from 1/5 corresponding to the 
asymptotic pion DA  (\ref{asymptotic}). Hence, if
\begin{equation}
  \langle \xi^2\rangle = \frac15 + \frac{12}{35} a_2
\label{relations2}
\end{equation} 
is determined with a given accuracy at some reference scale 
$\mu_0$ by a certain nonperturbative method, and $a_2$ is then obtained 
from the relation (\ref{relations2}), the error on $a_2$ is strongly 
amplified by the subtraction of the asymptotic contribution. This effect 
is well known and has been observed in all calculations up to date. 
The error on $a_2$ is relevant as it propagates through the 
renormalization group equations. In other words, although using 
$a_2$ as a nonperturbative parameter instead of $\langle \xi^2\rangle$ 
for the pion DA at a low reference scale $\phi_\pi(x,\mu^2_0)$ is just 
a rewriting, this choice is much more adequate in order to describe
the pion DA at high scales, $\phi_\pi(x,Q^2)$, $Q \gg \mu_0$, which 
enters QCD factorization theorems. Another issue to consider is that 
the relation in Eq.~(\ref{leibniz}) and therefore (\ref{relations1a}), 
(\ref{relations1b}), (\ref{relations2}) can be broken by lattice 
artifacts. Thus the choice of suitable operators requires some care. 
We will discuss our choice in more detail in the next section.

\section{Lattice formulation}

\label{sect:latform}

While the above relations refer to renormalized operators in Minkowski space,
we now move to Euclidean space and define the bare operators
\begin{align}
\cO^-_{\rho \mu \nu} (x)  
 & = \bar{d}(x) \left[ \Dl{(\mu} \Dl{\nu} 
  - 2 \Dl{(\mu} \Dr{\nu} + \Dr{(\mu} \Dr{\nu} \right] \gamma_{\rho)}  \gamma_5 \, u(x) \,, 
\nonumber \\
\cO^+_{\rho \mu \nu} (x)  
 & = \bar{d}(x) \left[ \Dl{(\mu} \Dl{\nu} 
  + 2 \Dl{(\mu} \Dr{\nu} 
  + \Dr{(\mu} \Dr{\nu} \right] \gamma_{\rho)} \gamma_5 \, u(x) 
\label{eq:opmultiplets}
\end{align} 
as our operator basis. On the lattice the covariant derivatives will be
replaced by their discretized versions. 

The operator $\cO^-_{\rho \mu \nu}$ can be written in a conventional 
shorthand notation as
\begin{align}
\cO^-_{\rho \mu \nu} (x)  
 & = \bar{d}(x) \Dd{(\mu} \Dd{\nu} \gamma_{\rho)} \gamma_5 \, u(x) 
\end{align}
and its matrix element between the vacuum and the pion state is proportional 
to the bare lattice value of 
$\langle (x - (1 - x))^2 \rangle = \langle \xi^2\rangle$.
In the continuum, the operator 
$\cO^+_{\rho \mu \nu}$ is the second derivative of the axial-vector current:
\begin{equation} \label{eq:plusrelation}
\cO^+_{\rho \mu \nu} (x) = 
\partial_{(\mu} \partial_\nu \cO_{\rho)} (x) \quad \mbox{with} \quad
\cO_\rho (x) = \bar{d}(x) \gamma_{\rho} \gamma_5 u(x) \,.
\end{equation} 
However, this relation is violated on the lattice because of 
discretization errors in the derivatives. The distinction 
between $\cO^+_{\rho \mu \nu}$ and 
$\partial_{(\mu} \partial_\nu \cO_{\rho)}$ for finite lattice spacing 
appears to be numerically important and will be discussed in detail in 
what follows. Note that $\cO^+$ is the Euclidean analogue of the 
Minkowski-space operator 
$\mathcal M^{(0,2)} + 2 \mathcal M^{(1,1)} + \mathcal M^{(2,0)}$
such that its matrix element between the vacuum and the pion state
corresponds to the bare value of 
$\langle (x + 1 - x)^2 \rangle = \langle 1^2 \rangle$.

The corresponding renormalized (e.g., in the $\MS$ scheme) 
axial-vector current is then given by
\begin{equation}
\cO^{\MS}_\rho (x) = Z_A \cO_\rho (x) 
\end{equation} 
with $Z_A \neq 1$ on the lattice. 

In order to express its matrix elements in terms of the physical
quantities introduced in Minkowski space we apply the rules
\begin{equation}
\gamma^0_{\mathrm M} = \gamma_4 \; , \; \gamma^j_{\mathrm M} = i \gamma_j 
\end{equation} 
for $j=1,2,3$, where the subscript M distinguishes the Minkowski objects.
Consequently, 
\begin{equation}
\gamma_5^{\mathrm M} = i \gamma^0_{\mathrm M} \gamma^1_{\mathrm M} 
 \gamma^2_{\mathrm M} \gamma^3_{\mathrm M} 
 = - \gamma_1 \gamma_2 \gamma_3 \gamma_4 = - \gamma_5 \,.
\end{equation} 
The components of the three-vector $\bf{p}$ of the spatial momentum
of the pion will be denoted by $p_j$, although they are equal to the
contravariant space components of the Minkowski momentum $p$. The time
component of the Minkowski momentum is identified with the corresponding 
energy: $p_0 = E_\pi ({\bf p})$. In this way one gets in Euclidean 
notation
\begin{align}
\langle 0 | \cO^{\MS}_4 (0) | \pi ({\bf p}) \rangle & =
- i E_\pi ({\bf p}) f_\pi \,, \\
\langle 0 | \cO^{\MS}_j (0) | \pi ({\bf p}) \rangle & =
- p_j f_\pi \,.
\end{align} 

Similarly, the Euclidean space components of the coordinate vector $x$
are identified with the contravariant components of the Minkowski 
space-time four-vector, while for the time components we have
$x_0 = -i x_4$. This entails the following rule for the covariant
derivatives:
\begin{equation}
-i D_0^{\mathrm M} = D_4 \; , \; D_j^{\mathrm M} = D_j \,.
\end{equation} 
Therefore we find, 
e.g., for $j \neq k$
\begin{equation} \label{eq:xi2me}
\langle 0 | \cO^{\MS -}_{4jk} ( 0) | \pi ({\bf p}) \rangle 
= i f_\pi \langle \xi^2 \rangle E_\pi ({\bf p}) p_j p_k \,.
\end{equation} 

The operators $\cO^-_{\rho \mu \nu}$ and $\cO^+_{\rho \mu \nu}$
mix under renormalization even in the continuum. On the lattice the
continuous rotational $O(4)$ symmetry of Euclidean space is broken and 
reduced to the discrete $H(4)$ symmetry of the hypercubic lattice. 
This symmetry breaking can introduce additional mixing operators. 
It can even lead to mixing of the operators of interest with 
operators of lower dimension such that the mixing coefficients  
are proportional to powers of $1/a$. This complicates the 
renormalization procedure significantly.
However, it may be possible to choose the lattice operators such that they
belong to an irreducible representation of $H(4)$ which forbids mixing 
with further operators, in particular with lower-dimensional operators. 
In the present case there is one such choice, given by the 
operators $\cO^\pm_{\rho \mu \nu}$ with all three indices different.
For the computation of the required matrix elements we can restrict
ourselves to the operators
(see, e.g., \cite{Braun:2006dg,Arthur:2010xf})
\begin{equation}\label{eq:operators}
  \mathcal O^{\pm}_{4jk}\,, \qquad j \neq k \in \{1,2,3\}\,.
\end{equation}
The renormalized operators are then given by
\begin{align}
\cO^{\MS -}_{4jk} (x) & = Z_{11} \cO^-_{4jk} (x)
 + Z_{12} \cO^+_{4jk} (x) \,,
\nonumber \\
\cO^{\MS +}_{4jk} (x) & = Z_{22} \cO^+_{4jk} (x) \,.
\end{align} 
Note that due to the discretization artifacts in the derivatives
one cannot expect $Z_{22}$ to be equal to $Z_A$. 

For the calculation of $\langle \xi^2 \rangle^{\MS}$ and $a_2^{\MS}$ we are
now left with two tasks: computation of the bare matrix elements and
evaluation of the renormalization factors. We extract the bare matrix 
elements from two-point correlation functions of the operators 
$\cO^\pm_{\rho\mu\nu}$ and $\cO_\rho$ with suitable interpolating
fields $J(x)$ for the $\pi$-mesons. For the latter we consider the two 
possibilities
\begin{align}
 J_5(x) &= \bar u(x)\gamma_5 d(x)\,,
\nonumber \\
 J_{45}(x) &= \bar u(x)\gamma_4\gamma_5 d(x) \label{eq:inter}
\end{align}
with smeared quark fields. The details of our smearing algorithm will be
given below. Let
\begin{align}
 C^A_{\rho}(t,{\bf p}) &=  a^3 \sum_{\bf x} e^{-i{\bf p\cdot x}} 
\langle \cO_\rho({\bf x},t) J_A(0) \rangle,
\nonumber \\
C^{\pm;A}_{\rho\mu\nu}(t,{\bf p}) &= a^3 \sum_{\bf x}e^{-i{\bf p\cdot x}}
\langle \cO^\pm_{\rho\mu\nu}({\bf x},t) J_A(0) \rangle,
\label{eq:corr_function} 
\end{align}
where $A=5$ or $A=45$, $\bf{p}$ is the three-vector of the spatial 
momentum, and the summation goes over the set of spatial lattice 
points ${\bf x}$ for a given Euclidean time $t$. 

For times $t$, where the correlation functions are saturated by the 
contribution of the lowest-mass pion state, we expect that, e.g.,   
\begin{widetext}
\begin{equation}
 C^{\pm;A}_{\rho\mu\nu}(t,{\bf p}) =
  \langle 0 | \cO^\pm_{\rho\mu\nu} (0)|\pi({\bf p})\rangle   
  \langle \pi({\bf p})|J_A(0) |0 \rangle \frac{1}{2E} 
  \left[e^{-Et} + \tau_{\mathcal O}\tau_{J}e^{-E(T - t)} \right] \,.
\end{equation}
\end{widetext}
Here $E \equiv E_\pi({\bf p})$, $T$ is the temporal extent of our 
lattice, and the $\tau$-factors take into account
transformation properties of the correlation functions under time reversal. 
One finds $\tau_{J_5} = -1$, $\tau_{J_{45}} = 1$, 
$\tau_{\cO} = 1$ for the operators $\cO^\pm_{4jk}$, $\cO_{4}$
and $\tau_{\cO} = -1$ for $\cO_j$, where $j,k=1,2,3$. 
We utilize these symmetries in order to reduce the statistical 
fluctuations of our raw data, i.e., we average over the
two corresponding times $t$ and $T-t$ with the appropriate sign factors.

From the ratios
\begin{equation} \label{eq:ratios-of-CFs}
 \mathcal R^{\pm;A}_{\rho\mu\nu; \sigma} =  
\frac{C^{\pm;A}_{\rho\mu\nu}(t,{\bf p})}{C^A_{\sigma}(t,{\bf p})}
\end{equation}
we can extract the required bare matrix elements 
$\langle 0 |\cO^\pm_{\rho\mu\nu} (0)|\pi({\bf p})\rangle$, which carry 
the information on the second moment of the pion DA.

Equation~(\ref{eq:xi2me}) shows that a calculation of matrix 
elements of $\cO^\pm_{4jk}$ requires two nonvanishing spatial components 
of the momentum. We choose them as small as possible, $p = 2\pi/L$, 
where $L$ is the spatial extent of our lattice. To suppress 
statistical fluctuations we average over the possible directions, e.g., 
${\bf p} = (p,p,0)$, ${\bf p} = (p,-p,0)$, ${\bf p} = (-p,p,0)$, 
${\bf p} = (-p,-p,0)$ for $j=1$, $k=2$.
If the correlation functions are dominated by the single-pion states,
the time-dependent factors in the ratios of correlation functions 
cancel and we obtain, e.g., for the operator $\cO^\pm_{412}$ and 
the momentum ${\bf p} = (p,p,0)$
\begin{equation}
\mathcal R^{\pm;A}_{412; 4} = - \left(\frac{2\pi}{L}\right)^2 R^\pm \,, 
\end{equation}
where the constants $R^\pm$ are related to the bare 
lattice values of the second moment of the pion DA through
\begin{align}
  \langle \xi^2 \rangle^{\mathrm {bare}} = R^-,  &&
   a_2^{\mathrm {bare}} =  \frac{7}{12}\left( 5 R^- - R^+\right)\,.
\end{align}
They should not depend on the choice of the interpolating field $J_A$.
Note that $R^+ \slashed{=} 1$ and therefore for bare quantities
\begin{align}\label{eq:ineq}
    a_2^{\mathrm {bare}} &\slashed{=} \frac{7}{12}\left(5 \langle \xi^2 \rangle^{\mathrm {bare}} - 1\right)\,.
\end{align}
For the renormalized moments in the $\MS$ scheme we obtain
\begin{align}
 \langle \xi^2 \rangle^{\MS} &= \zeta_{11} R^- 
+  \zeta_{12} R^+ \,,
\nonumber \\
  a_2^{\MS} &= \frac{7}{12} \Big[ 5 \zeta_{11} R^- 
  + \big(5 \zeta_{12} -  \zeta_{22}\big) R^+\Big] \,,\label{eq:ren_moments}
\end{align}
where 
\begin{equation} 
\zeta_{11} = \frac{Z_{11}}{Z_A} \; , \;
\zeta_{12} = \frac{Z_{12}}{Z_A} \; , \;
\zeta_{22} = \frac{Z_{22}}{Z_A}
\end{equation}
are ratios of renormalization constants defined in the next section.

In the continuum limit we expect that
\begin{align} 
Z_{22} \langle 0 | \cO^+_{4jk}(0) | \pi ({\bf p}) \rangle 
 &= - Z_A p_j p_k \langle 0 | \cO_4 (0) | \pi ({\bf p}) \rangle 
\notag\\ 
&= i p_j p_k E_\pi ({\bf p}) f_\pi \,.
\end{align}
Hence the quantity 
\begin{equation}\label{eq:I2} 
\langle 1^2 \rangle^{\MS} := \frac{Z_{22}}{Z_A}
\frac{\langle 0 | \cO^+_{4jk}(0) | \pi ({\bf p}) \rangle}
 {(-p_j p_k) \langle 0 | \cO_4 (0) | \pi ({\bf p}) \rangle } 
= \zeta_{22} R^+
\end{equation}
should approach unity as the lattice spacing tends to zero. In this case
the relation 
\begin{equation} 
a_2^{\MS} = \frac{7}{12} \big( 5  \langle \xi^2 \rangle^{\MS} -1\big) 
\label{a2-from-xi2}
\end{equation} 
is recovered (cf.\ Eq.~(\ref{relations1b})),
whereas for finite lattice spacing it follows from (\ref{eq:ren_moments})
\begin{equation} 
a_2^{\MS} = \frac{7}{12} \big( 5  \langle \xi^2 \rangle^{\MS} 
                             - \langle 1^2 \rangle^{\MS}\big)\,. 
\label{a2-from-xi2a}
\end{equation} 
We emphasize that Eq.~(\ref{a2-from-xi2}) is only recovered in the 
continuum limit, which is always delicate. There are two possibilities: Either 
$\langle \xi^2\rangle$ is measured on the lattice, the result extrapolated 
to zero lattice spacing, and at the final step $a_2$ is obtained using 
the relation (\ref{a2-from-xi2}), or $a_2$ is calculated directly on the 
lattice and then extrapolated to the continuum limit. The first approach 
was used in Refs.~\cite{Braun:2006dg,Arthur:2010xf} whereas in this work 
we use the second method.

\section{Renormalization constants}
\label{sect:renco}

\begin{table} 
\renewcommand{\arraystretch}{1.2}
\caption{Ensembles used for nonperturbative renormalization.} 
\label{table:reno-ens}
\begin{ruledtabular}
\begin{tabular}{ccc}
$\beta$  &  $\kappa$  & Size    \\ \hline
5.20     &  0.13550   & $32^3 \times 64$  \\  
5.20     &  0.13584   & $32^3 \times 64$  \\  
5.20     &  0.13596   & $32^3 \times 64$  \\ \hline 
5.29     &  0.13620   & $32^3 \times 64$  \\  
5.29     &  0.13632   & $32^3 \times 64$  \\  
5.29     &  0.13640   & $64^3 \times 64$  \\ \hline 
5.40     &  0.13640   & $32^3 \times 64$  \\  
5.40     &  0.13647   & $32^3 \times 64$  \\  
5.40     &  0.13660   & $48^3 \times 64$    
\end{tabular}
\end{ruledtabular}
\renewcommand{\arraystretch}{1.0}
\end{table}

From our bare matrix elements we have to compute the corresponding 
renormalized matrix elements in the $\MS$ scheme, which is used in 
the perturbative calculations. In the continuum we therefore have to 
deal with the renormalization of the two mixing operator multiplets 
given in Eq.~(\ref{eq:opmultiplets}). Note that $\cO^+_{\rho \mu \nu}$, 
being the second derivative of the axial-vector current, has vanishing 
forward matrix elements, at least in the continuum.

On the lattice we work with the operator multiplets 
\begin{equation} \label{eq:multi1}
\cO^+_{423} \; , \; \cO^+_{413} \; , \; \cO^+_{412} \; , \; \cO^+_{123} 
\end{equation} 
and
\begin{equation} \label{eq:multi2}
\cO^-_{423} \; , \; \cO^-_{413} \; , \; \cO^-_{412} \; , \; \cO^-_{123} \,. 
\end{equation} 
Under the hypercubic group $H(4)$, both multiplets transform identically 
according to a four-dimensional irreducible 
representation~\cite{Gockeler:1996mu}. The symmetry
properties of these multiplets ensure that they do not mix with any other
operators. Because of the well-known shortcomings of lattice perturbation 
theory we want to determine the renormalization and mixing factors 
nonperturbatively on the lattice, utilizing a variant of the RI'-MOM scheme. 
However, since forward matrix elements of $\cO^+_{\rho \mu \nu}$ 
eventually vanish, we cannot use the momentum geometry of the original
RI'-MOM scheme but have to work with a kind of RI'-SMOM 
scheme~\cite{Sturm:2009kb}.

In order to describe our renormalization procedure we consider a somewhat
more general situation than what is needed in this paper. Let 
$\cO^{(m)}_i (x)$ ($i=1,2,\ldots,d$, $m=1,2,\ldots,M$) denote $M$ 
multiplets of local quark-antiquark operators which transform 
identically according to some irreducible, unitary, $d$-dimensional 
representation of $H(4)$. Call the unrenormalized, but (lattice-)regularized
vertex functions (in the Landau gauge) $V^{(m)}_i (p,q)$, where $p$ 
and $q$ are the external quark momenta. The corresponding renormalized
(in the $\MS$ scheme) vertex functions are denoted by $\bar{V}^{(m)}_i (p,q)$.
The dependence of $\bar{V}^{(m)}_i$ on the renormalization scale $\mu$ 
is suppressed for brevity. Note that $V^{(m)}_i$ carries Dirac indices 
and is therefore to be considered as a $4 \times 4$-matrix. (The color 
indices have been averaged over.)

We choose 
\begin{equation} 
p = \frac {\mu}{\sqrt{2}} (1,1,0,0) \; , \; 
q = \frac {\mu}{\sqrt{2}} (0,1,1,0) 
\end{equation}
such that $p^2 = q^2 = (p-q)^2 = \mu^2$. As our renormalization condition
we take (in the chiral limit)
\begin{eqnarray} 
\lefteqn{\sum_{i=1}^d \mathrm {tr} 
    \left( \hat{B}^{(m)}_i \hat{B}^{(m') \dagger}_i \right)=}
\nonumber\\
 &=& Z_q^{-1}  \sum_{m''=1}^M \hat{Z}_{m m''} \sum_{i=1}^d 
   \mathrm {tr} \left( V^{(m'')}_i \hat{B}^{(m') \dagger}_i \right),
\end{eqnarray}
where $\hat{B}^{(m)}_i$ is the lattice Born term corresponding to $V^{(m)}_i$.
The wave function renormalization constant of the quark fields $Z_q$
is determined from the quark propagator, as usual~\cite{Gockeler:2010yr}, 
and subsequently converted to the $\MS$ scheme. Using the lattice Born 
term instead of the continuum Born term and proceeding analogously in the 
calculation of $Z_q$ ensures that $\hat{Z}$ is the unit matrix in the 
free case.

The renormalization matrix $\hat{Z}$ leads from the bare operators on
the lattice to renormalized operators in our SMOM scheme. The matrix
$Z$ transforming the bare operators into renormalized operators in the
$\MS$ scheme is then given by $Z = C \hat{Z}$, where the 
matrix $C$ is defined as
\begin{equation} 
\sum_{m''=1}^M \sum_{i=1}^d C_{m m''} \mathrm {tr} 
    \left( B^{(m'')}_i B^{(m') \dagger}_i \right)
 = \sum_{i=1}^d \mathrm {tr} 
   \left( \bar{V}^{(m)}_i B^{(m') \dagger}_i \right) .
\end{equation}
Here $\bar{V}^{(m)}_i$ is the renormalized vertex function in the
$\MS$ scheme and $B^{(m)}_i$ is the continuum Born term such that
the conversion matrix $C$ is completely determined from a continuum 
calculation.

\begin{table} 
\renewcommand{\arraystretch}{1.2}
\caption{\label{table:fit_choices}Choices for the fits.} 
\label{table:fits}
\begin{ruledtabular}
\begin{tabular}{cccccc}
Fit    & Fit interval & $n_{\mathrm {loops}}$ & Lattice   &  $r_0$   & $r_0 \Lambda_{\MS}$ \\
number & (in GeV$^2$)  &                 & artifacts & (in fm)  &                  \\ \hline
 1 & $4 \, < \mu^2 < 100 \,$ & 2 & 
 $A_3 \neq 0$ & $0.50 \, $ & 0.789 \\
 2 & $2 \,  < \mu^2 < 30 \, $ & 2 & 
 $A_3 \neq 0$ & $0.50 \, $ & 0.789 \\
 3 & $4 \, < \mu^2 < 100 \, $ & 1 & 
 $A_3 \neq 0$ & $0.50 \,$ & 0.789 \\
 4 & $4 \, < \mu^2 < 100 \, $ & 2 & 
 $A_3=0$ & $0.50 \,$ & 0.789 \\
 5 & $4 \, < \mu^2 < 100 \,$ & 2 & 
 $A_3 \neq 0$ & $0.49 \,$ & 0.789 \\
 6 & $4 \, < \mu^2 < 100 \, $ & 2 & 
 $A_3 \neq 0$ & $0.50 \,$ & 0.737
\end{tabular}
\end{ruledtabular}
\renewcommand{\arraystretch}{1.0}
\end{table}

\begin{table} 
\renewcommand{\arraystretch}{1.2}
\caption{Fit results at $\beta=5.40$ for $\mu_0^2 = 4 \, \mathrm {GeV}^2$.} 
\label{table:fitresults}
\begin{ruledtabular}
\begin{tabular}{ccccccc}
 {} & Fit 1 & Fit 2 & Fit 3 & Fit 4 & Fit 5 & Fit 6 \\ \hline
$\zeta_{11}$ & 2.026 & 2.031 & 2.123 & 2.001 & 2.040 & 2.041 \\
$\zeta_{12}$ 
  & $-0.199$ & $-0.205$ & $-0.233$ & $-0.188$ & $-0.202$ & $-0.203$ \\
$\zeta_{22}$ & 1.474 & 1.476 & 1.479 & 1.467 & 1.474 & 1.474 
\end{tabular}
\end{ruledtabular}
\renewcommand{\arraystretch}{1.0}
\end{table}

Here we have to consider the cases $M=2$, $d=4$ for the multiplets 
(\ref{eq:multi1}), (\ref{eq:multi2}) and $M=1$, $d=4$ for the 
axial-vector current. The required $\MS$ vertex functions in the 
chiral limit for up to two loops can be extracted from 
Refs.~\cite{Gracey:2011fb,Gracey:2011zg}. As we are only interested 
in ratios of renormalization factors, $Z_q$ drops out and is not needed. 
In the following we describe our method for the determination
of the renormalization matrix of the multiplets (\ref{eq:multi1}), 
(\ref{eq:multi2}). The procedure for the ratios with $Z_A$ is completely
analogous, because the anomalous dimension of the nonsinglet axial-vector
current vanishes.

The calculation of the vertex functions with the help of momentum 
sources is straightforward. Partially twisted boundary conditions 
applied to the quark propagators allow us to vary the renormalization 
scale $\mu$ independently of the lattice size. 
The ensembles used for the evaluation of the $Z$ matrices according to
the above formulas are listed in Table~\ref{table:reno-ens}. Due
to the rather small quark masses the subsequent chiral extrapolation 
appears to be quite safe.
  
Ideally, the renormalization scale $\mu$ should satisfy the conditions
\begin{equation} 
  1/L^2 \ll \Lambda^2_{\mathrm {QCD}} \ll \mu^2 \ll 1/a^2 
\end{equation}
for a lattice with lattice spacing $a$ and extent $L$. Then 
lattice artifacts would be negligible and the scale dependence
could be described by low-order continuum perturbation theory. However,
the above conditions are hard to realize in practice and the $Z$-values 
at any given scale suffer from discretization artifacts as well as from
truncation errors of the perturbative expansions. Therefore we try to 
exploit as much of the available nonperturbative information as possible
by performing a joint fit of the $\mu$-dependence of the chirally 
extrapolated renormalization matrices $Z(a,\mu)_{\mathrm {MC}}$ 
for our three $\beta$-values $\beta = 5.20$, $5.29$ and $5.40$. 

The choice of the fitting procedure is motivated by the following 
considerations. The (perturbative) running of the $Z$-matrices is 
governed by the anomalous dimension matrix
\begin{equation} 
\gamma = - \left( \mu \frac{d Z}{d \mu} \right) Z^{-1} \,.
\end{equation}
Introducing the running renormalized coupling $g(\mu)$ with 
$\mu \, dg/d\mu = \beta(g)$ we get 
\begin{equation} 
\frac{d Z}{d g} = - \frac{\gamma (g)}{\beta (g)} Z \,.
\end{equation}
This system of differential equations can formally be solved in the form
\begin{widetext}
\begin{equation} 
Z(\mu) Z^{-1} (\mu_0) = 
\sum_{n=0}^\infty (-1)^n \int_{g(\mu_0)}^{g(\mu)} dg_n 
  \int_{g(\mu_0)}^{g_n} dg_{n-1} \cdots 
  \int_{g(\mu_0)}^{g_2} dg_1 \frac{\gamma(g_n)}{\beta(g_n)} \cdots 
   \frac{\gamma(g_2)}{\beta(g_2)} \frac{\gamma(g_1)}{\beta(g_1)} \,.
\end{equation}
\end{widetext}
\begin{figure*}
\includegraphics*[width=8.0cm]{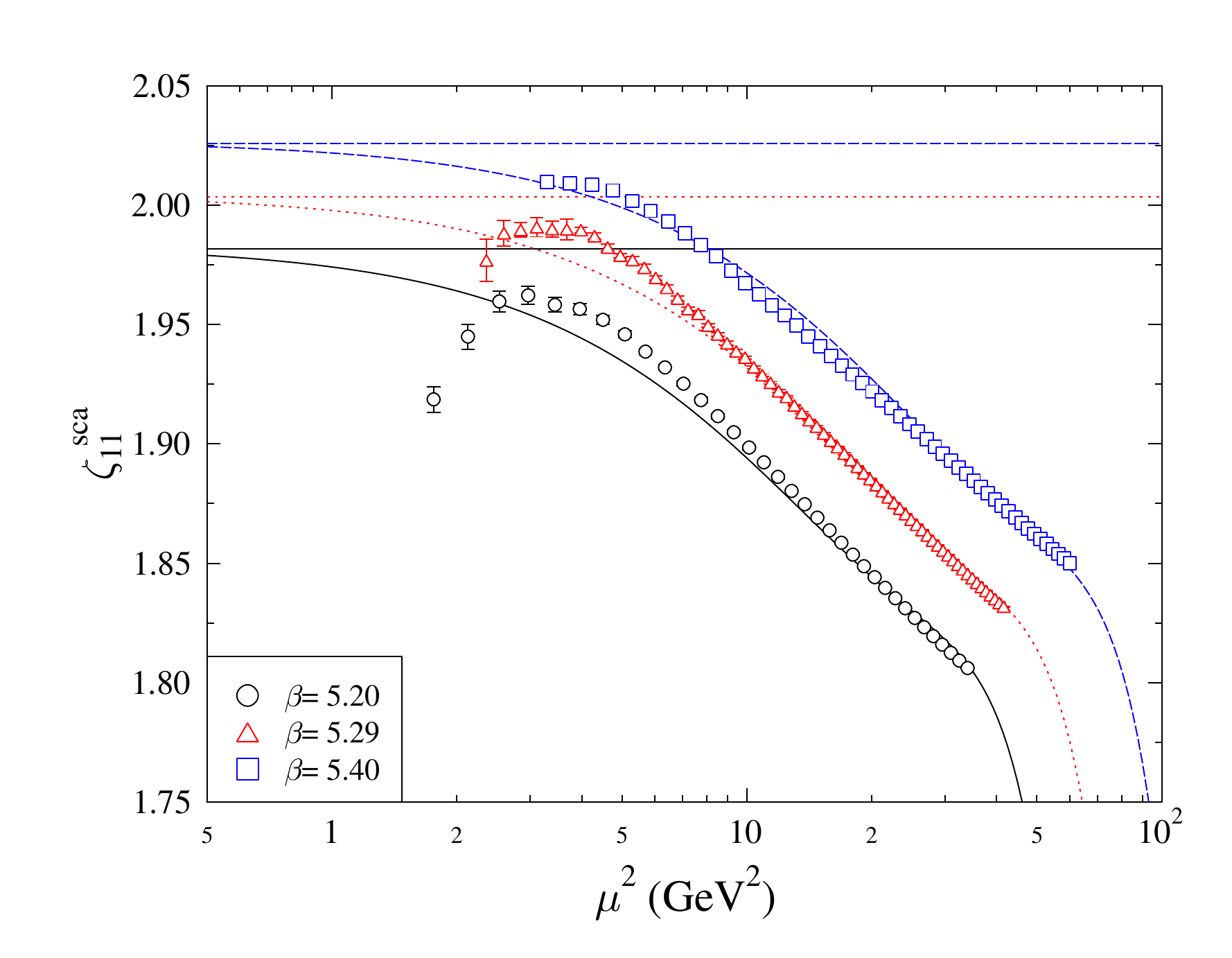}
\includegraphics*[width=8.0cm]{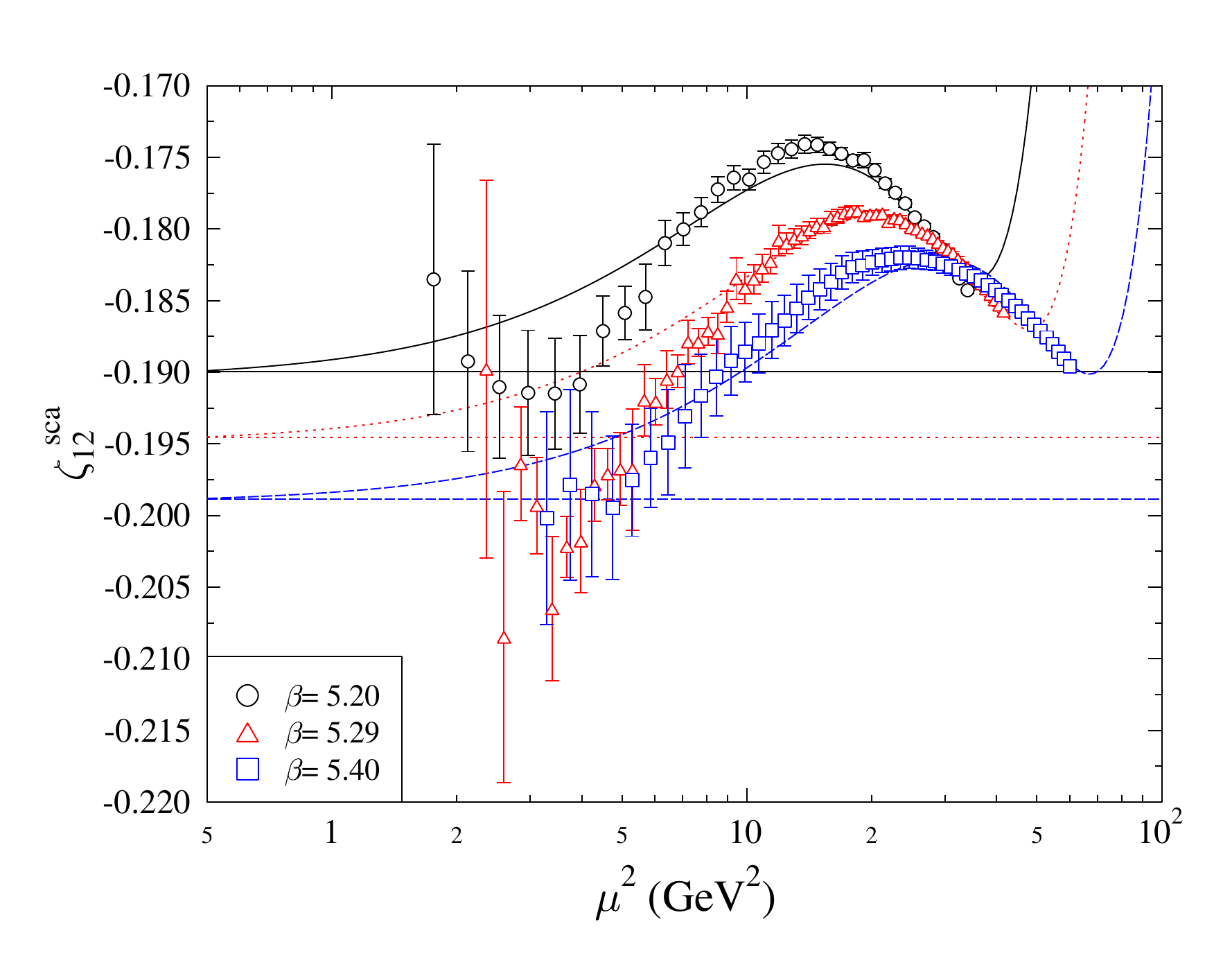} \\
\includegraphics*[width=8.0cm]{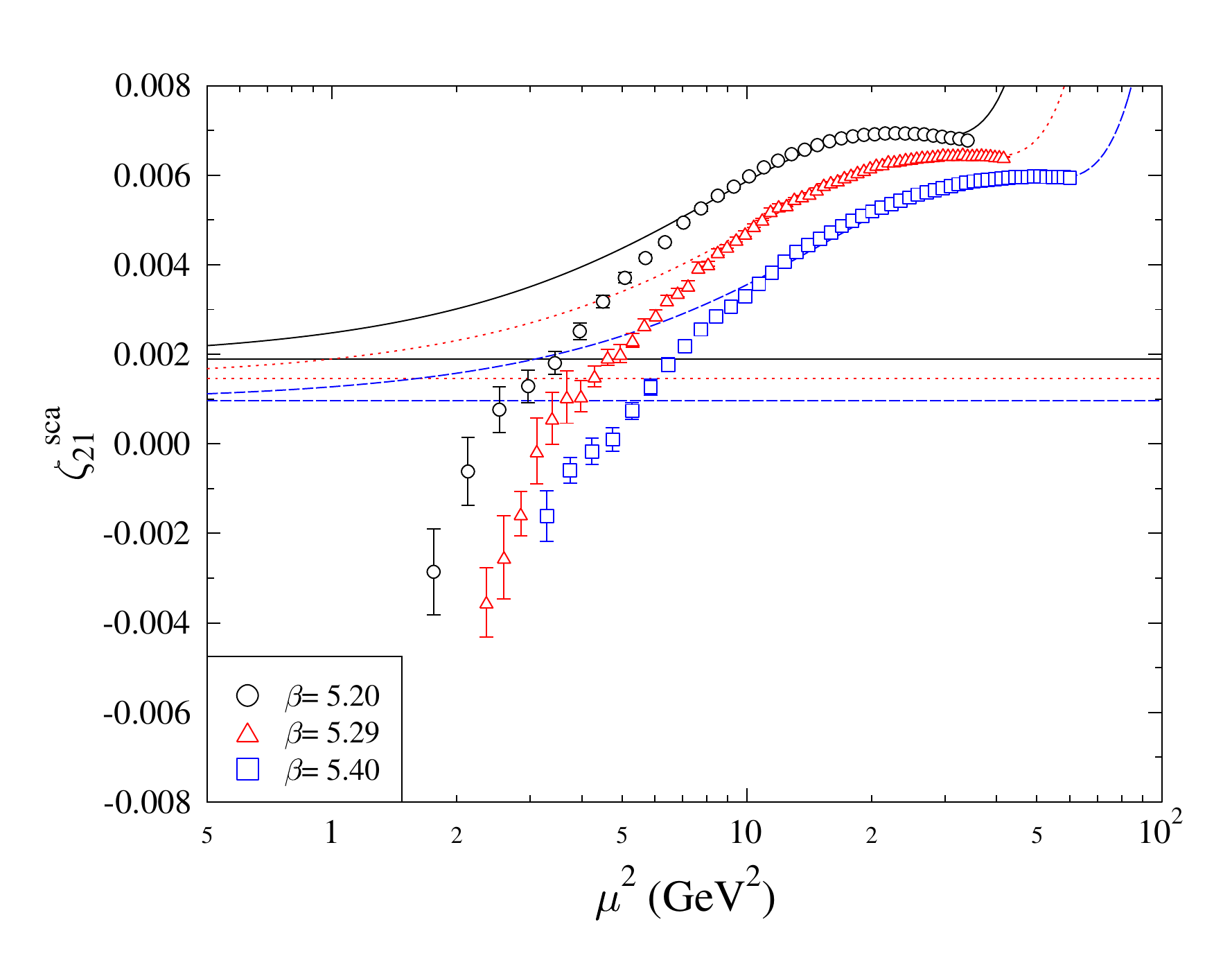}
\includegraphics*[width=8.0cm]{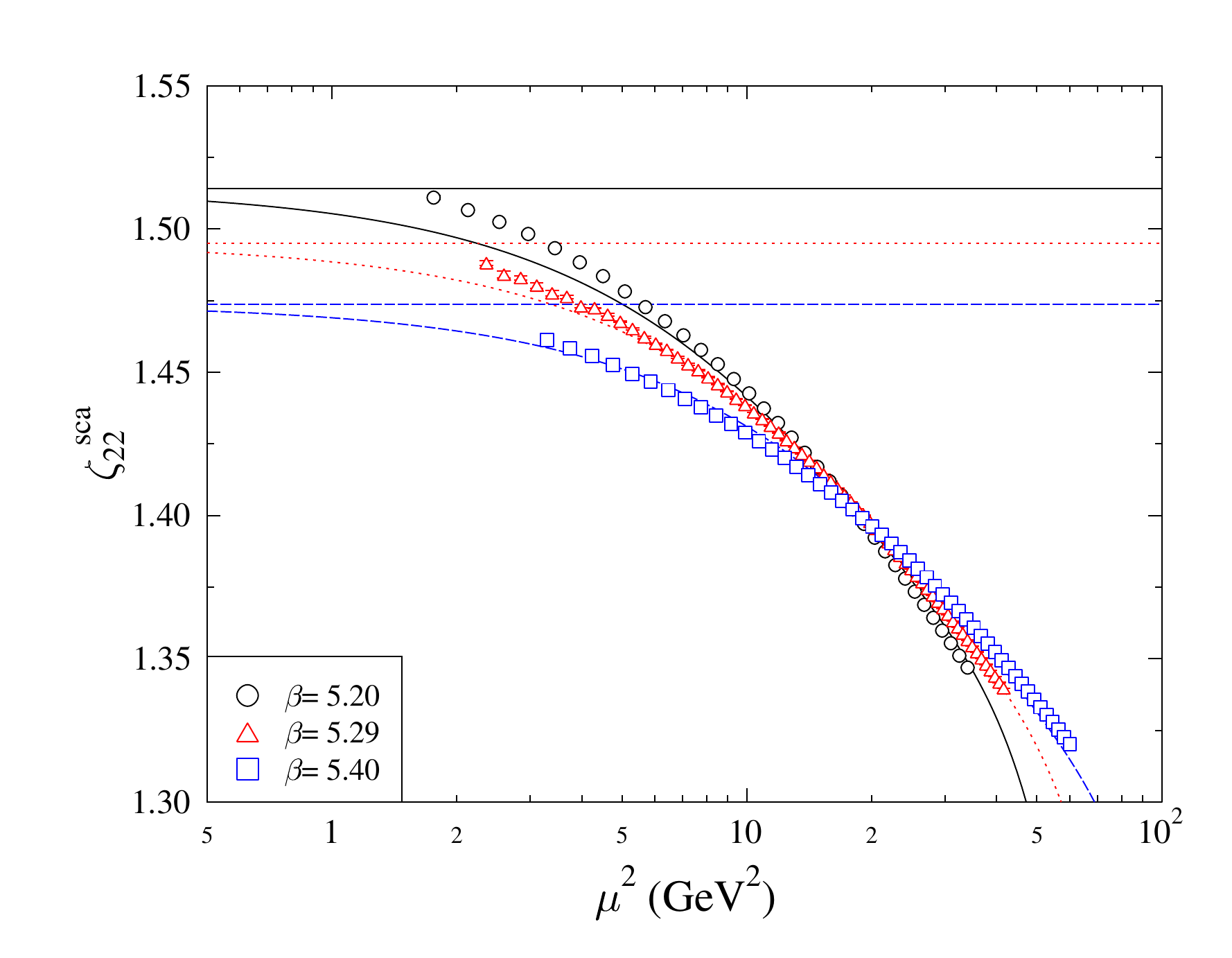}
\caption{Renormalization and mixing factors $\zeta_{ij}$
in the chiral limit, perturbatively scaled to $\mu_0 = 2 \, \mathrm {GeV}$
(cf.\ Eq.~(\ref{eq:figs})) together with curves representing Fit 1. 
The error bars show the statistical errors. The horizontal lines
indicate the fitted values $\zeta_{ij} (a, \mu_0)$.
Note that the fit is aimed at describing the data for large values
of the scale $\mu$, the fit interval being 
$4 \, \mathrm {GeV}^2 < \mu^2 < 100 \, \mathrm {GeV}^2$.}
\label{fig:z}
\end{figure*}
From the three-loop anomalous dimension matrix one can calculate a
corresponding approximation of $W(\mu, \mu_0):= Z(\mu) Z^{-1} (\mu_0)$,
which should describe the $\mu$-dependence for sufficiently large scales
$\mu$ if there were no discretization effects. Adding a plausible 
ansatz for an effective description of these lattice artifacts we arrive
at the following fit function for the matrices $Z(a,\mu)_{\mathrm {MC}}$:
\begin{align} \label{eq:fitfun}
Z(a,\mu)_{\mathrm {MC}} &= W(\mu, \mu_0) Z(a,\mu_0) + A_1 a^2 \mu^2 
  + A_2 (a^2 \mu^2)^2
\notag\\
&{} + A_3 (a^2 \mu^2)^3 \,.
\end{align}
The fit parameters are the entries of the three renormalization matrices
$Z(a,\mu_0)$ at the reference scale $\mu_0$ and the entries of the three
matrices $A_i$ parameterizing the lattice artifacts. 
Note that we allow for a nonvanishing value of $Z_{21}$ although $Z_{21}$
vanishes in the continuum. 

The statistical errors of the data are quite small, in particular for 
larger scales, and the resulting statistical errors of the fit
parameters turn out to be unrealistically tiny. Therefore the
statistical errors will be
ignored in the following. The systematic uncertainties, on the other 
hand, are much more important. In order to estimate them we perform a 
number of fits varying exactly one element of the analysis at a time.
More precisely, we choose as representative examples for fit intervals 
$4 \, \mathrm {GeV}^2 < \mu^2 < 100 \, \mathrm {GeV}^2$ and 
$2 \, \mathrm {GeV}^2 < \mu^2 < 30 \, \mathrm {GeV}^2$, and we use
the expressions for the $\MS$ vertex functions $\bar{V}^{(m)}_i$
with $n_{\mathrm {loops}}= 1,2$. For the parameterization of the lattice 
artifacts we either take the complete expression in Eq.~(\ref{eq:fitfun})
or we set $A_3=0$. Finally, we consider values for $r_0$ and 
$r_0 \Lambda_{\MS}$ corresponding to the results given in 
Ref.~\cite{Fritzsch:2012wq}. The various possibilities are compiled
in Table~\ref{table:fits}.
 
As an example we show the fit results for $\beta=5.40$ in 
Table~\ref{table:fitresults}, choosing $\mu_0^2 = 4 \, \mathrm {GeV}^2$.
The numbers for the other $\beta$-values are similar.

The largest effect comes from the variation of $n_{\mathrm {loops}}$:
Working with the 1-loop vertex functions increases the result for
$\zeta_{11}$ by about 5\%, and the modulus of the mixing coefficient 
$\zeta_{12}$ increases even by about 17\%. In order to obtain our
final numbers for $\langle \xi^2 \rangle^{\MS}$, $a_2^{\MS}$ and 
$\langle 1^2 \rangle^{\MS}$
we extract them from the raw data for $R^\pm$ using each of these
sets of values for $\zeta_{11}$, $\zeta_{12}$ and $\zeta_{22}$.    
So we get six results for each of our gauge field ensembles.
As our central values we take the results from Fit 1. 
Defining $\delta_i$ as the difference between
the result obtained with the $\zeta$s from Fit $i$ and the result 
determined with the $\zeta$s from Fit $1$, we estimate the  
systematic uncertainties due to the renormalization factors 
as $\sqrt{\delta_2^2 + (0.5 \cdot \delta_3)^2 + \delta_4^2 + 
          \delta_5^2 + \delta_6^2}$. Here we have multiplied $\delta_3$
by $1/2$, because going from two loops to three or more loops in the 
perturbative vertex functions is expected to lead to a smaller change than
going from one loop to two loops. This should amount to a rather 
conservative error estimate.

In Fig.~\ref{fig:z} we show the entries of the matrix
\begin{align} 
 W^{-1}(\mu, \mu_0) Z(a,\mu)_{\mathrm {MC}} &= 
 Z(a,\mu_0) + W^{-1}(\mu, \mu_0)
\notag\\ &{}\hspace*{-1cm} \times \Big[ A_1 a^2 \mu^2  +  A_2 (a^2 \mu^2)^2 +  A_3 (a^2 \mu^2)^3 \Big] \label{eq:figs}
\end{align}
for $\mu_0^2 = 4 \, \mathrm {GeV}^2$ at our three $\beta$-values along 
with the fit curves resulting from Fit 1 in Table~\ref{table:fits}. 
The horizontal lines represent the fitted values $\zeta_{11} (a,\mu_0)$ etc.

In the previous paper~\cite{Braun:2006dg} the renormalization and mixing 
factors were evaluated in a mixed perturbative-nonperturbative approach, 
based on the representation of $\cO^+_{\rho \mu \nu}$ as the second 
derivative of the axial-vector current (see Eq.~(\ref{eq:plusrelation})).
Repeating this calculation in a completely nonperturbative setting
we find that the overall renormalization factor corresponding to 
$\zeta_{11}$ agrees within a few percent. The nonperturbative 
mixing coefficient, on the other hand, has the same (negative) sign 
as its perturbatively computed counterpart, but its modulus is up
to one order of magnitude larger. This observation underlines the 
necessity of nonperturbative renormalization, at least for the presently
reachable $\beta$-values.

\section{Analysis of the bare data}
\label{sect:anadata}

As was already mentioned in Sec.~\ref{sect:latform}, the bare matrix 
elements related to the pion DA's second moments can be extracted 
from ratios of lattice correlation functions given by 
Eq.~(\ref{eq:ratios-of-CFs}). We briefly describe our procedure. 

The gauge field configurations used in this work have been generated 
with the Wilson gauge action and $n_f=2$ flavors of nonperturbatively 
improved Wilson fermions. We have analyzed ${\mathrm O}(1000 - 2000)$
configurations for three different values of the gauge coupling,
$\beta = 5.20,5.29,5.40$, and pion masses in the range 
$m_\pi \sim 500 - 150 \, \mathrm {MeV}$. The lattice spacings 
and spatial volumes 
vary between $0.06 - 0.081$ fm and $(1.71 - 4.57 \, \mathrm{fm})^3$, 
respectively. A list of our ensembles can be found 
in Table~\ref{table:ListOfLattices}. For scale setting we
used the Sommer parameter with the value $r_0 = 0.5$ fm
~\cite{Bali:2012qs, Fritzsch:2012wq}.

\begin{table}
\renewcommand{\arraystretch}{1.2}
\caption{Ensembles used for this work.} \label{table:ListOfLattices}
\begin{ruledtabular}
\begin{tabular}{lcccr}
\multicolumn{1}{c}{$\kappa$} & $m_\pi /$ MeV & Size & $m_\pi L$ & Number of\\
 & & & & configs.\footnote{The number of measurements per configuration is shown in parentheses. \\ \hspace*{-0.5cm}
$^\dagger$These ensembles were generated on the QPACE systems, financed 
primarily by the SFB/TR 55, while the others were generated earlier within the QCDSF collaboration.}\\
\hline \multicolumn{5}{c}{$\beta = 5.20, a = 0.081\ \text{fm}, a^{-1} = 2400\ \text{MeV}$} \\
\hline 0.13596$^\dagger$ & 280 & $32^3 \times 64$ & 3.7 & $1999 (\times 4)$ \\
\hline \multicolumn{5}{c}{$\beta = 5.29, a = 0.071\ \text{fm}, a^{-1} = 2800\ \text{MeV}$} \\
\hline 0.13620$^\dagger$ & 430 & $24^3 \times 48$ & 3.7 & $1764 (\times 2)$ \\
0.13620$^\dagger$ & 422 & $32^3 \times 64$ & 4.8 & $1998 (\times 2)$\\
0.13632 & 294 & $32^3 \times 64$ & 3.4 & $1999 (\times 1)$ \\
0.13632 & 289 & $40^3 \times 64$ & 4.2 & $2028 (\times 2)$ \\
0.13632$^\dagger$ & 285 & $64^3 \times 64$ & 6.7 & $  1237 (\times 2)$ \\
0.13640$^\dagger$ & 150 & $64^3 \times 64$ & 3.5 & $1599 (\times 3)$ \\
\hline \multicolumn{5}{c}{$\beta = 5.40, a = 0.060\ \text{fm}, a^{-1} = 3300\ \text{MeV}$} \\
\hline 0.13640 & 491 & $32^3 \times 64$ & 4.8 & $982 (\times 2)$ \\
0.13647$^\dagger$ & 430 & $32^3 \times 64$ & 4.2 & $1999 (\times 2)$ \\
0.13660 & 260 & $48^3 \times 64$ & 3.8 & $2178 (\times 2)$
\end{tabular}
\end{ruledtabular}
\renewcommand{\arraystretch}{1.0}
\end{table}

The correlation functions (\ref{eq:corr_function}) have been computed
for the operators $\cO_4, \cO_{4jk}$ (see Eq.~(\ref{eq:operators})) 
leading to the ratios $\mathcal R^{\pm;J_{\mathrm {opt}}}_{4 jk; 4} $, 
where $J_{\mathrm {opt}}$ is discussed below. On most of the ensembles, 
we performed more than one measurement per configuration to increase the 
statistics. The source positions for the correlation functions were 
selected randomly to reduce the autocorrelations among 
configurations lying close to one another in the Monte Carlo history.  
We want the interpolating operators to have a good
overlap with the ground state of the pion. To this end, 
Wuppertal smearing~\cite{Gusken:1989ad,Gusken:1989qx}
was applied to the sources, with APE smeared~\cite{Falcioni:1984ei} 
gauge fields.

In order to reduce the overlap with excited states even further
we have used the variational method~\cite{Michael:1982gb, 
Michael:1985ne, Luscher:1990ck, Blossier:2009kd} 
with the two interpolators (\ref{eq:inter}) to obtain an optimal 
interpolator $J_{\mathrm {opt}} = \alpha J_{5} + \beta J_{45}$. 
This procedure is based on the $t$-dependent $2 \times 2$-matrix of 
two-point correlation functions of the interpolating fields $J_5$ 
and $J_{45}$, projected onto vanishing spatial momentum. Solving a 
generalized eigenvalue problem for this matrix allows one to determine
$J_{\mathrm {opt}}$ from the eigenvector belonging to the lowest 
energy eigenvalue. Using this interpolator in the correlation 
functions improves the signal of the ground state. We have also 
tried to apply the Additional Interpolators Method with a third, 
time-shifted interpolator \cite{Schiel:2015kwa,Aubin:2010jc}, but the results 
changed only marginally. Our final numbers will be based on the 
results obtained with $J_{\mathrm {opt}}$. This differs from the 
approach of Ref.~\cite{Braun:2006dg}, where only the interpolator $J_5$
was utilized in the final analysis.

\begin{figure*}
\includegraphics*[width=8.0cm]{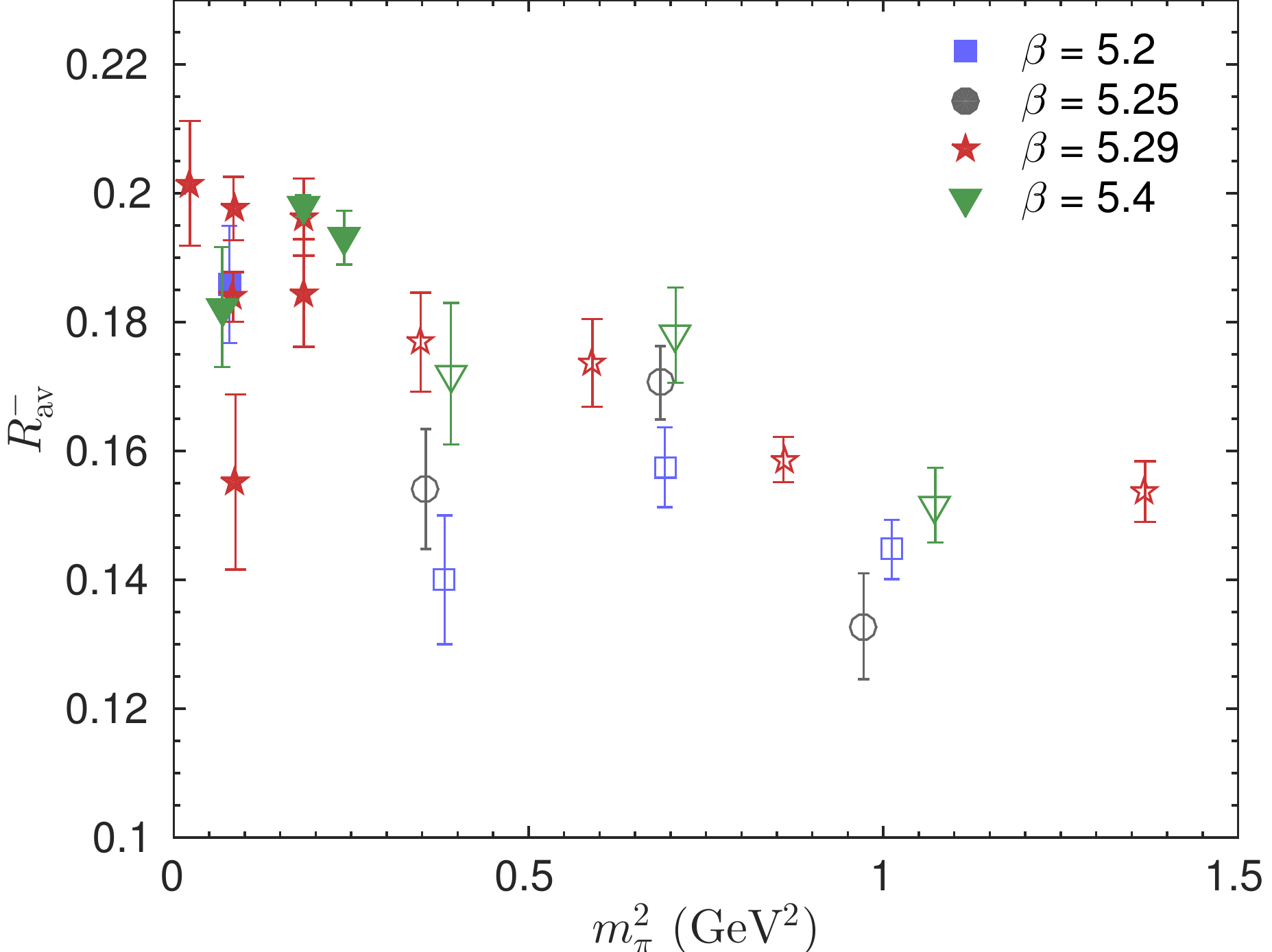}
\includegraphics*[width=8.0cm]{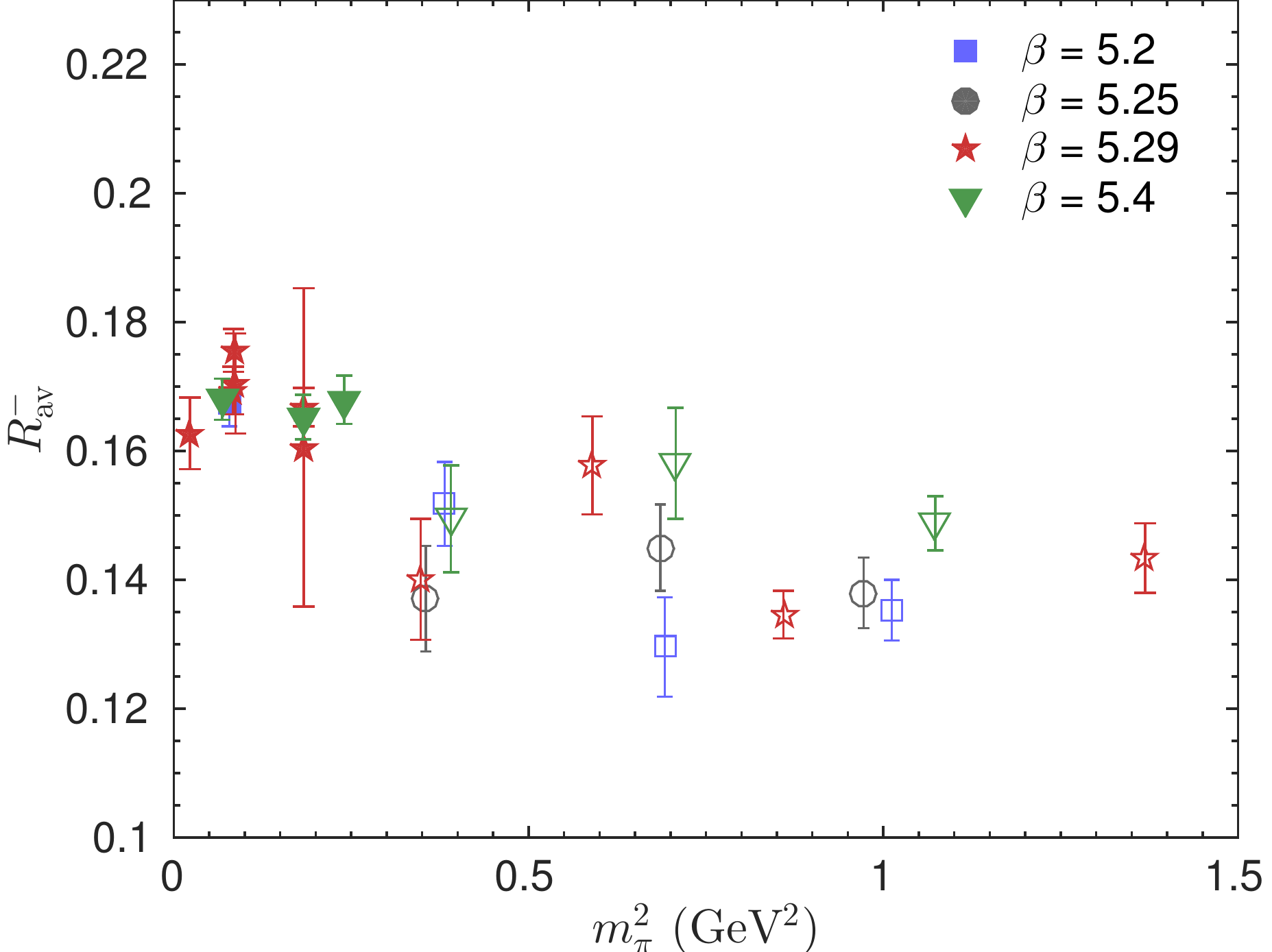}
\caption{\label{fig:all_data} Bare results for $R_{\mathrm {av}}^-$ 
from this work (filled symbols) and from \cite{Braun:2006dg} (open symbols)
for the two interpolators $J_{45}$ (left panel) and $J_5$ (right panel).}
\end{figure*}

To suppress statistical fluctuations, we have averaged over 
all possible values of $j,k$, and all possible momentum directions, 
\begin{equation}
   R^{\pm}_{\mathrm {av}} = \left(\frac{L}{2\pi}\right)^2 \frac{1}{12}
   \sum_{j}\sum_{k> j} \sum_{p_j=\pm p}\sum_{p_k=\pm p}
   |\mathcal R^{\pm;J_{\mathrm {opt}}}_{4jk;4}|,
\end{equation}
where  $p = 2\pi /L$. The quantities $R_{\mathrm {av}}^\pm$ have then
been fitted to a constant in a time interval where a plateau could 
be identified. The choice of the fit ranges was based on the goodness
of the correlated $\chi^2$-values and the stability of the results
upon reducing the fit interval.
The statistical errors were evaluated using the Jackknife procedure 
combined with the binning method. We have observed that a 
binsize $n_{\mathrm {bin}} = 4$ saturates the statistical error, 
which means that the autocorrelations are satisfactorily taken into account.  

Our bare results are collected in Tables~\ref{table:results_R-_R+_J5},
\ref{table:results_R-_R+_J45} and \ref{table:results_Rpm} in the Appendix. 
In Fig.~\ref{fig:all_data} we display 
$R_{\mathrm {av}}^- = \langle \xi^2 \rangle^{\mathrm {bare}}$ 
for the two interpolating operators 
$J_{45}$ and $J_5$ together with the corresponding results obtained in 
Ref.~\cite{Braun:2006dg}. 
We observe that our data are consistent with the 
measurements in~\cite{Braun:2006dg}, but extend to
considerably smaller pion masses all the way down to the physical value. 
Nevertheless, in the next section we will see that 
taking into account Eq.~(\ref{eq:ineq}) and using the nonperturbatively
computed value of $\zeta_{12}$ leads to a significant shift in the 
final result.

\section{Renormalized results}
\label{sect:results}

In this section we present our results for the renormalized quantities
$\langle 1^2 \rangle^{\MS}$ (cf.~Eq.~(\ref{eq:I2})), 
$\langle \xi^2\rangle^\MS$ and $a_2^\MS$ 
(cf.~Eq.~(\ref{eq:ren_moments})). For each ensemble, the 
final error budget has to encompass the statistical errors coming 
from the determination of the bare quantities on the lattice, 
the systematic uncertainties due to the choice of the fit range, 
and the errors of the renormalization constants. 
The ensuing extrapolation to the physical pion mass and eventually 
to the continuum will introduce further uncertainties.
In order to include the errors coming from the renormalization constants 
we proceed as already indicated at the end of Sec.~\ref{sect:renco}. 
For every fit choice in Table~\ref{table:fit_choices}, 
we use the renormalization factors $\zeta_{11}$, $\zeta_{12}$, $\zeta_{22}$
resulting from this fit to compute the renormalized quantities from the 
bare ratios $R^\pm_{\mathrm {av}}$ according to 
Eqs.~(\ref{eq:ren_moments}) and (\ref{eq:I2}),
taking the correlations between $R^+_{\mathrm {av}}$ and 
$R^{-}_{\mathrm {av}}$ into account. The central
value is then taken from the first fit choice, and the error due to the 
renormalization constants is determined from the differences 
with the other fit choices, as described in Sec.~\ref{sect:renco}.
In the following plots we show the central values 
together with their statistical errors, while the errors 
coming from the renormalization constants are not included, but are 
given in the Tables.

\begin{figure}
\includegraphics[width=7.0cm]{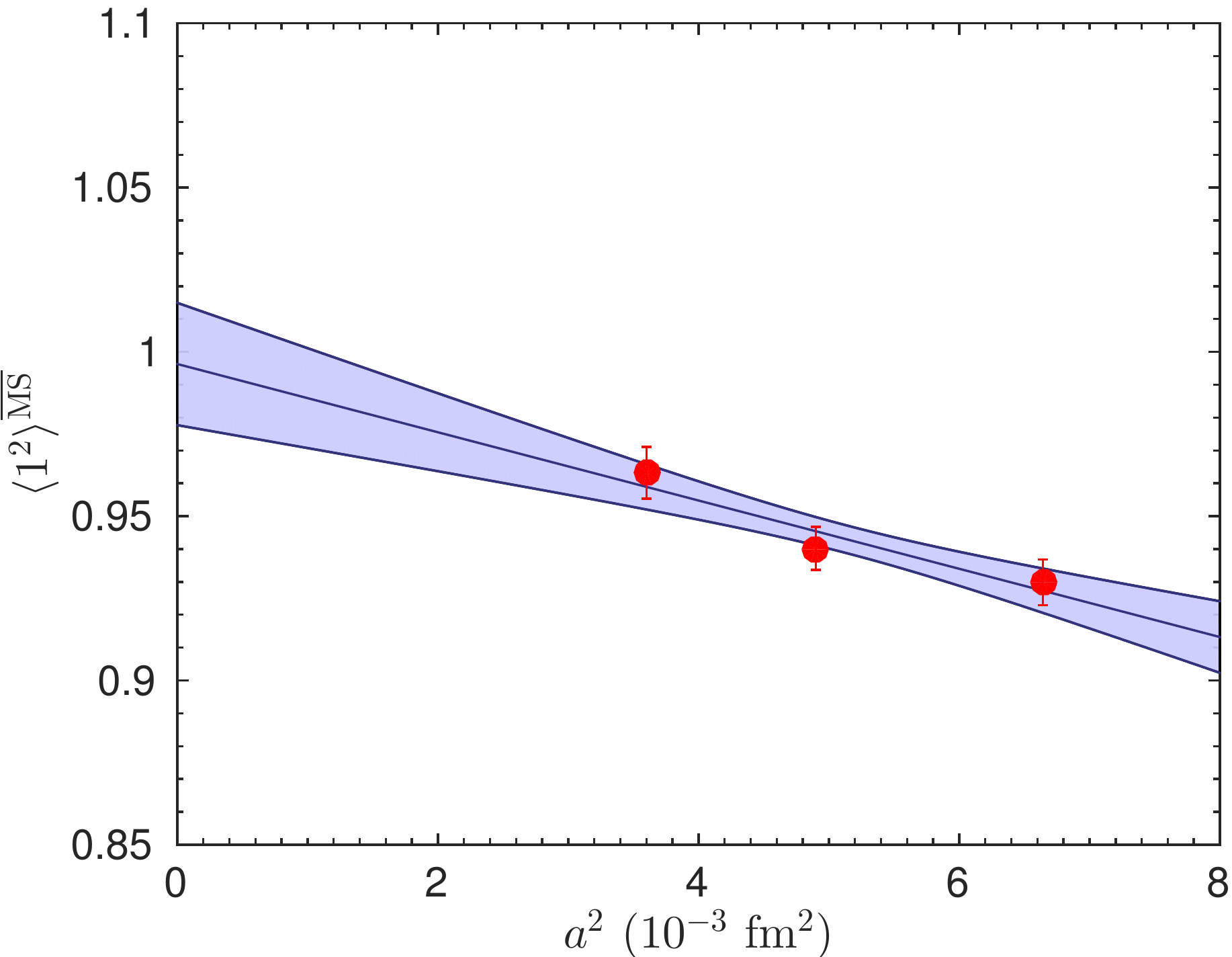}
\caption{ $\langle 1^2 \rangle^{\MS}$ as a function of the lattice 
spacing $a$ for ensembles with $m_\pi L \sim 3.4 -3.8$ and 
$m_\pi \sim 280 \, \mathrm {MeV}$. Only statistical errors are shown.}
\label{fig:I2}
\end{figure}

We start by presenting our results for $\langle 1^2 \rangle^{\MS}$. 
In the continuum limit, this quantity should be one for all pion masses.
Results for all ensembles are presented in Table~\ref{table:I2}.
In Fig.~\ref{fig:I2}, $\langle 1^2 \rangle^{\MS}$ is plotted for the three 
available lattice spacings using data for $m_\pi L \sim 3.4 - 3.8$ 
and $m_\pi \sim 260 - 294 \, \mathrm {MeV}$ 
(or $m_\pi \sim 280 \, \mathrm {MeV}$ for short; the mass dependence
is rather weak).

We also show an extrapolation to the continuum limit assuming a
linear dependence on $a^2$. 
We see that the result is consistent with unity within errors:
\begin{align}
\langle 1^2 \rangle^{\MS}_{a\to 0} &= 0.9963(186)(51)\,.
\end{align}
Here the first error is statistical, and the second error accounts
for the uncertainty due to the renormalization factors, estimated
as described at the end of Sec.~\ref{sect:renco}. 
It might be surprising that an extrapolation linear in $a^2$ works
so well although our operators are not $O(a)$-improved. However,
the covariant derivatives in the operator $\cO^+_{4jk}$ do not 
introduce $O(a)$ lattice artifacts, at least at tree level, and the 
$O(a)$ artifacts in $\cO_4$ should cancel to some extent between the 
numerator and the denominator in the ratio~(\ref{eq:I2}).
An extrapolation linear in $a$ looks less stable due
to the rather small range of $a$-values and yields a result which is 
a few percent larger.

Note that for $a^2 \sim 5\cdot 10^{-3}$~fm$^2$ corresponding to 
$\beta = 5.29$, where most of our data are collected, we obtain, e.g.,
at $m_\pi = 294 \, \mathrm {MeV}$ on a $32^3 \times 64$-lattice
\begin{align}
\langle 1^2 \rangle^{\MS}_{a \sim 0.07~\text{fm}} &= 0.9402(66)(54) \,.
\end{align}
The deviation from unity is only 6\%, however, it results in a $25-30\%$
increase in the value of $a_2^\MS$ at the same lattice spacing, calculated 
using Eq.~(\ref{a2-from-xi2a}) instead of the continuum relation
in Eq.~(\ref{a2-from-xi2}).   

The results for $\langle \xi^2\rangle^{\MS}$ and $a_2^\MS$ are
given in Tables~\ref{table:xi2} and \ref{table:a2}, where the first 
error is statistical and the second comes from the uncertainty 
in the determination of the renormalization constants. 
Ideally, one would now take the infinite volume limit, perform the 
continuum extrapolation at fixed pion masses and finally extrapolate
to the physical mass, if it is not included in the range of simulated 
masses. Unfortunately, our present set of data does not allow us to 
perform all three extrapolations in a controlled way. 

\begin{figure} 
\includegraphics[width=6.0cm]{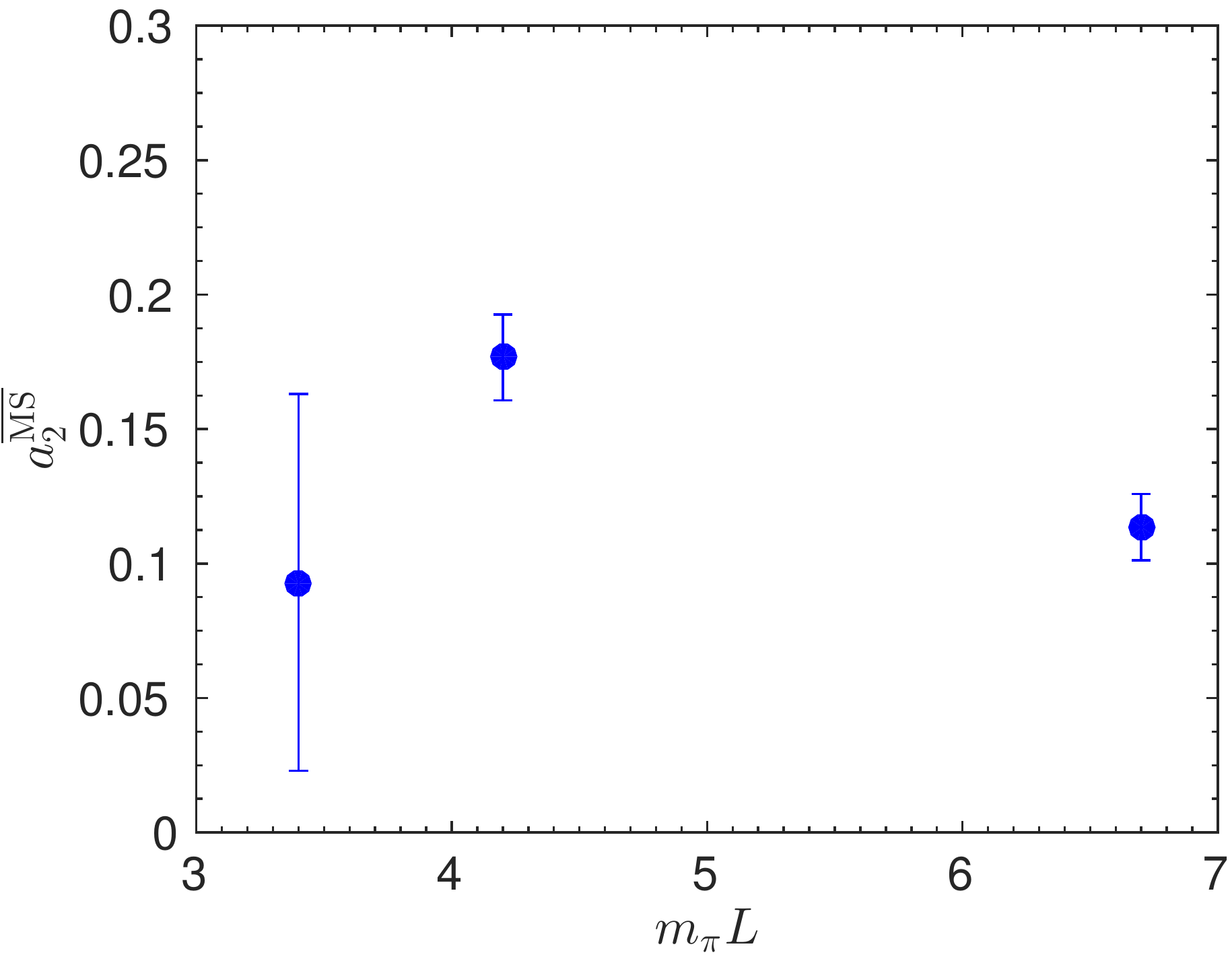} \\
\includegraphics[width=6.0cm]{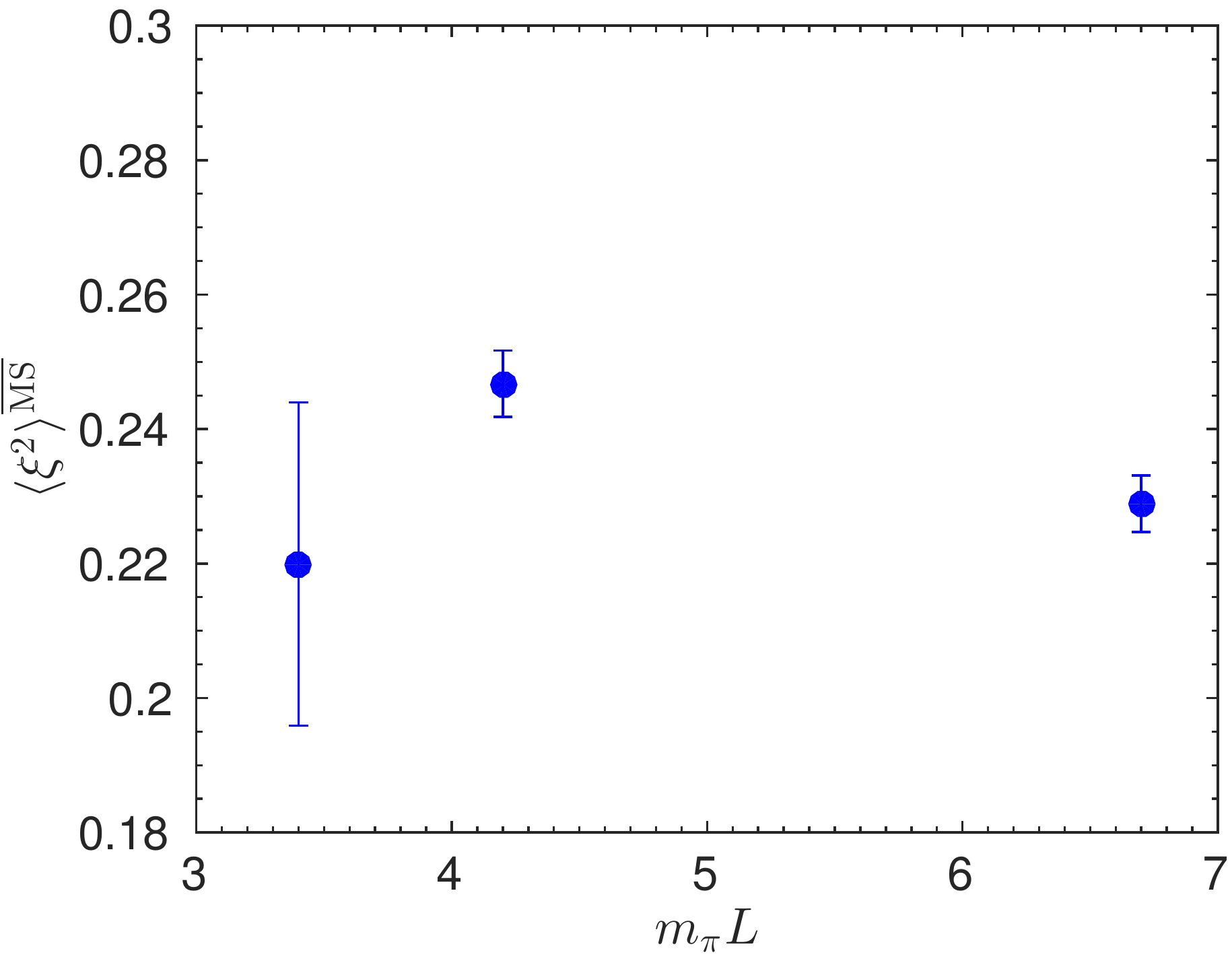} 
\caption{\label{fig:vol_effects} Renormalized results $a_2^{\MS}$
(upper panel) and $\langle \xi^2\rangle^{\MS}$ (lower panel)
as a function of $m_\pi L$ for ensembles with $\beta=5.29$ and  
$m_\pi \sim 290$ MeV. Only statistical errors are shown.}
\end{figure}

We can however study the finite size effects using the data at
$\beta = 5.29$, $\kappa = 0.13620$  ($m_\pi\sim 425 \, \mathrm {MeV}$)
and $\kappa = 0.13632$ ($m_\pi\sim 290 \, \mathrm {MeV}$), 
where we have two and three volumes, respectively. 
In Fig.~\ref{fig:vol_effects} we plot $a_2^{\MS}$ and 
$\langle \xi^2\rangle^{\MS}$ versus $m_\pi L$ 
for $m_\pi\sim 290 \, \mathrm {MeV}$ and see that there are 
indications of nonnegligible effects. In leading order chiral 
perturbation theory, on the other hand, there are no finite volume 
correction terms, as follows from the results in Ref.~\cite{Chen:2005js}.

\begin{figure}
\includegraphics[width=6.0cm]{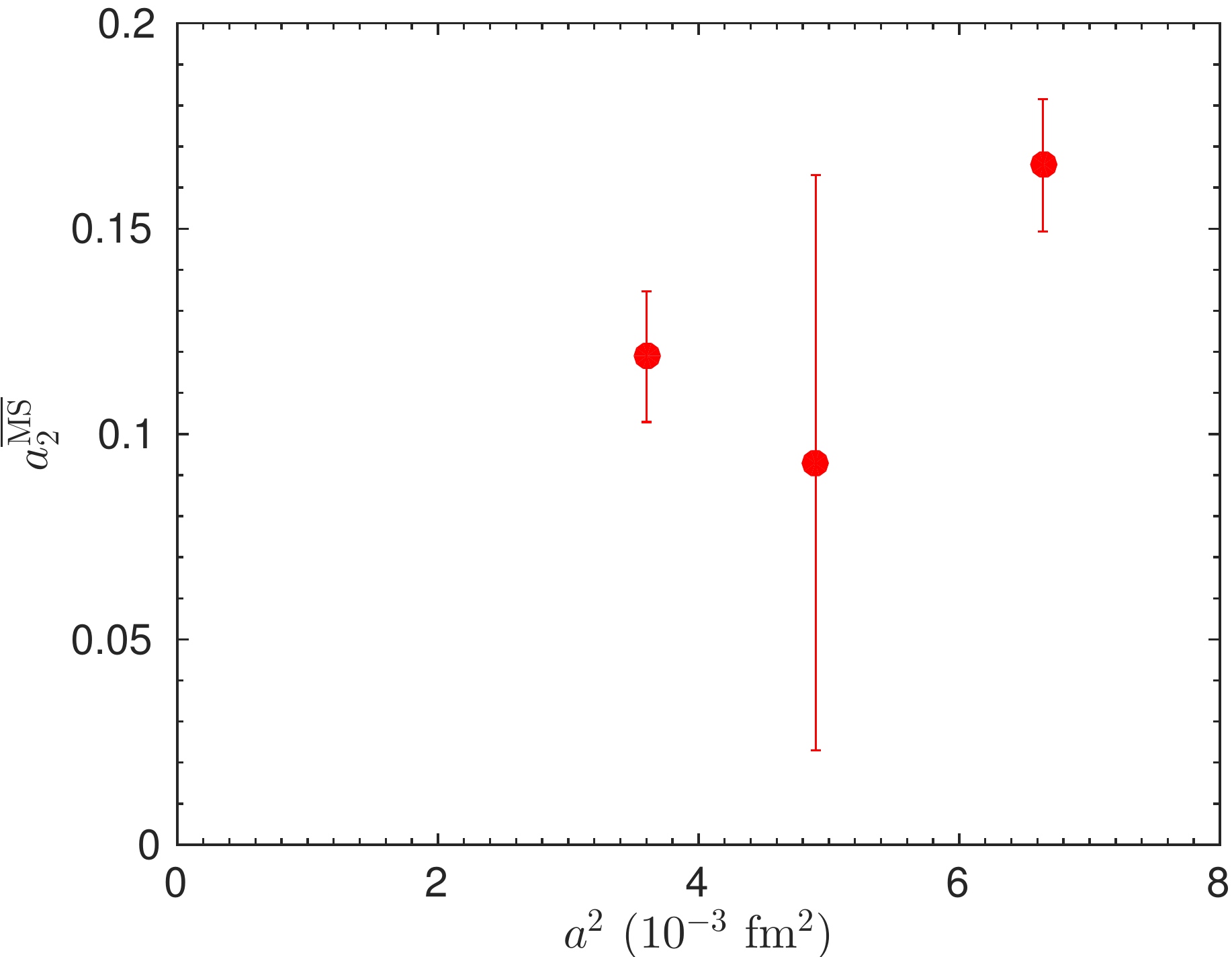} \\
\includegraphics*[width=6.0cm]{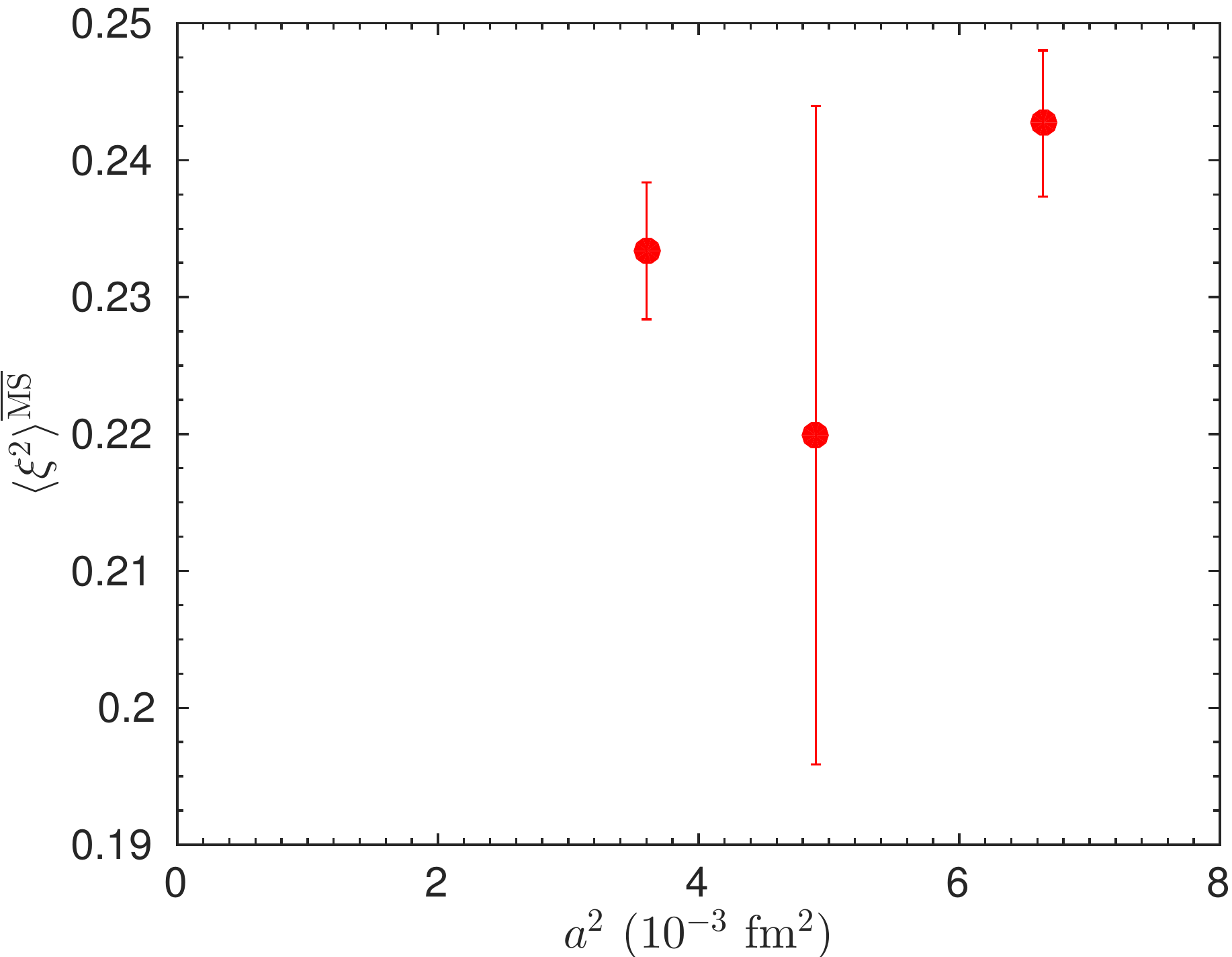}
\caption{\label{fig:contlim} Lattice spacing dependence of
$a_2^\MS$ (upper panel) and $\langle\xi^2\rangle^\MS$ (lower panel) for 
$m_\pi \sim 280\, \mathrm {MeV}$ and $m_\pi L \sim 3.4 - 3.8$. 
Only statistical errors are shown.}
\end{figure}

Similarly, we use our ensembles at $m_\pi \sim 280 \, \mathrm {MeV}$ 
and $m_\pi \sim 425\, \mathrm {MeV}$, where we have three 
and two different lattice spacings, respectively, to study 
discretization effects. Results for $a_2^{\MS}$ and 
$\langle\xi^2\rangle^\MS$ are shown in 
Fig.~\ref{fig:contlim}. Unfortunately, with only three lattice 
spacings at hand and relatively large statistical errors, it is 
impossible to perform a reliable continuum extrapolation. 

\begin{figure*}
\includegraphics*[width=8.0cm]{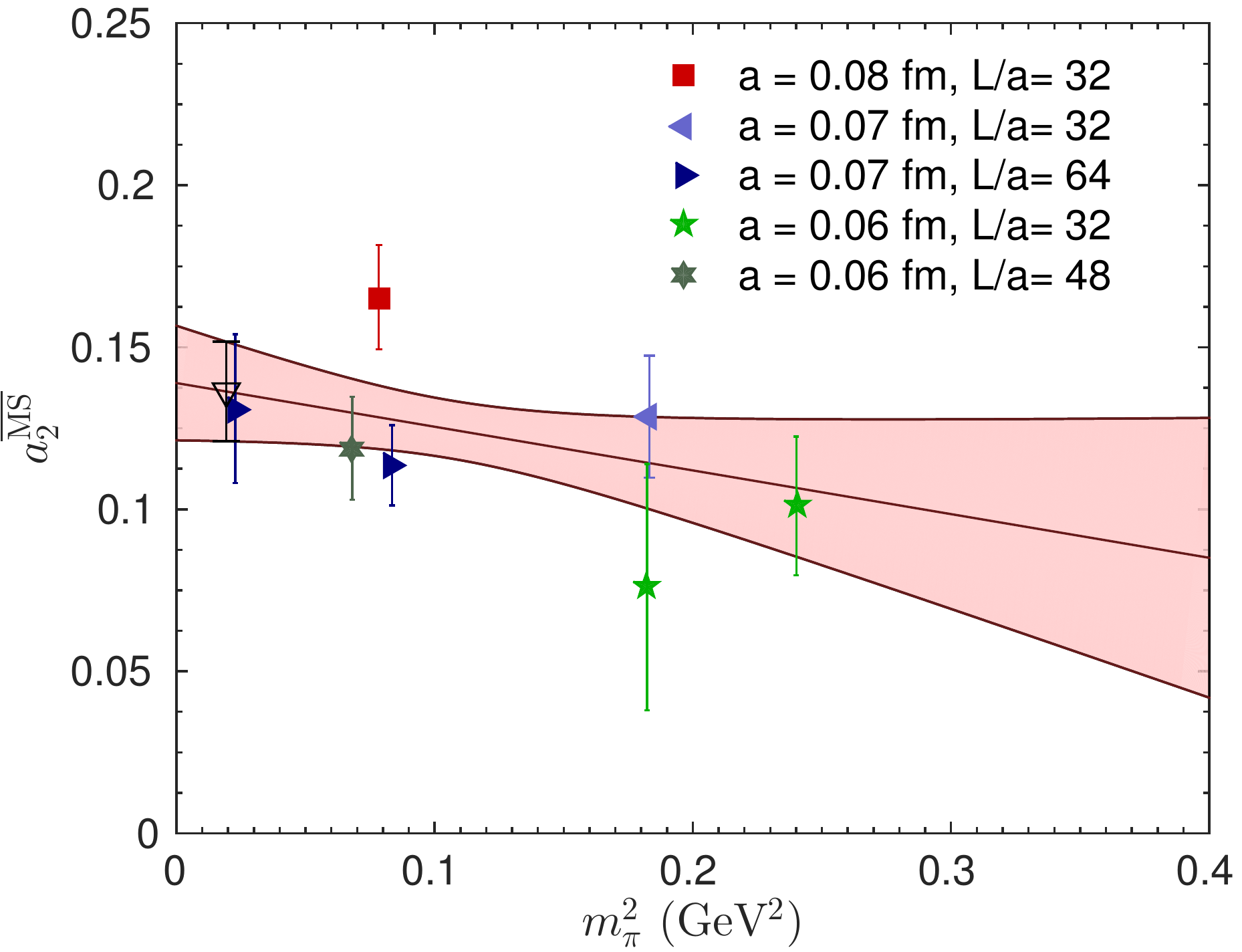}
\includegraphics*[width=8.0cm]{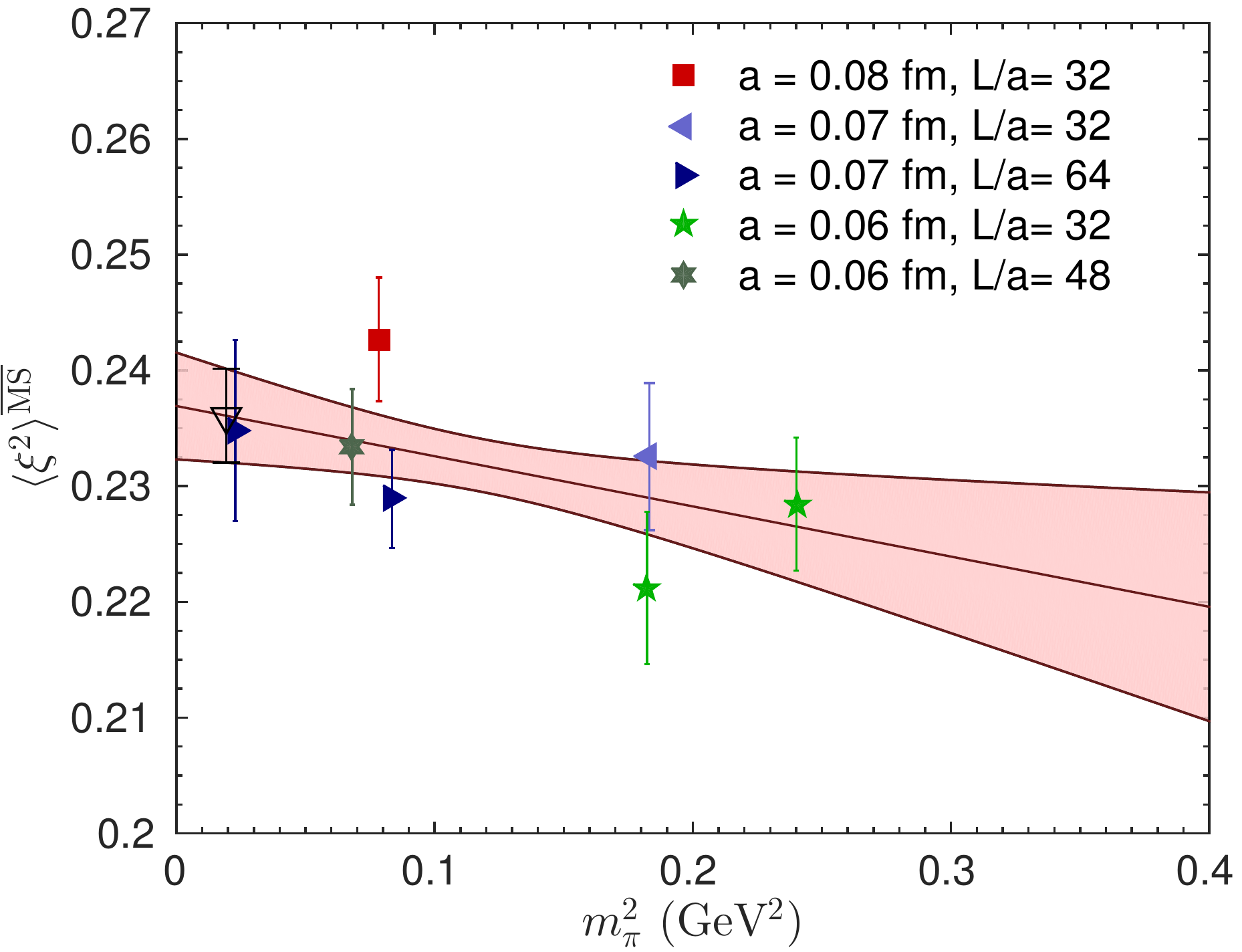}
\caption{\label{fig:chiral} Extrapolation to the physical
pion mass for $a_2^\MS$ (left panel) and $\langle\xi^2\rangle^\MS$ 
(right panel). The open triangle represents the extrapolated value.
Only statistical errors are shown.}
\end{figure*}

According to Ref.~\cite{Chen:2005js}, $\langle \xi^2 \rangle^{\MS}$, and 
hence also $a_2^{\MS}$, do not contain chiral logarithms, at least 
to one-loop order. Therefore we assume a linear dependence 
on $m_\pi^2$ for the extrapolation in the pion mass to the physical 
value. Since the ensemble with the lightest pion is already 
very close to the physical point, the chiral extrapolation is reliable.
As our lattice spacings do not vary that much,
and a proper continuum extrapolation of $\langle \xi^2 \rangle^{\MS}$ 
and $a_2^{\MS}$ cannot be attempted, we include results from all 
lattice spacings, but take into account only the data for the 
largest volume, where different volumes are available.
The resulting extrapolations of $a_2^{\MS}$ and
$\langle \xi^2 \rangle^\MS$ to the physical pion mass 
are plotted in Fig.~\ref{fig:chiral}. 
As in these fits $\chi^2/\mathrm {dof}$ is greater 
than one, we follow the procedure advocated by the Particle
Data Group~\cite{Agashe:2014kda} and multiply the errors by 
$\sqrt{\chi^2/\mathrm {dof}}$. As before, errors coming from 
the renormalization constants are not included in the plot. We 
perform an extrapolation for every fit choice given in 
Table~\ref{table:fit_choices} and compute the error of the final
number caused by the uncertainties of the renormalization factors
from the differences of the extrapolated results as indicated at 
the end of Sec.~\ref{sect:renco}.

From this procedure we find our final results
\begin{eqnarray}
   \langle \xi^2\rangle^\MS &=& 0.2361(41)(39)\,,
\nonumber\\
   a_2^\MS &=& 0.1364(154)(145) 
\label{finalnumbers}
\end{eqnarray}
at the scale $\mu = 2 \, \mathrm {GeV}$.
They can be compared with the earlier lattice calculations
\begin{align}
    \langle \xi^2\rangle^\MS &=0.269(39)\,,\quad a_2^\MS = 0.201(114)\,, 
& \quad \text{\cite{Braun:2006dg}}
\nonumber\\
   \langle \xi^2\rangle^\MS  &=0.28(1)(2)\,,\quad a_2^\MS = 0.233(29)(58)\,, 
& \quad \text{\cite{Arthur:2010xf}}
\end{align}
where, for \cite{Arthur:2010xf}, we have quoted the result 
for $\langle \xi^2\rangle^\MS $ on their larger lattice and used the 
continuum relation in Eq.~(\ref{a2-from-xi2}) to 
calculate the corresponding value of the second Gegenbauer moment $a_2^\MS$.

It should, however, be kept in mind that all these numbers were obtained
on lattices with lattice spacings between $0.06$ and $0.08 \, \mathrm {fm}$. 
The investigation of discretization effects for $\langle \xi^2\rangle^\MS$ 
and $a_2^{\MS}$ will remain a challenge for future studies.

\section{Conclusions and Outlook}

We have presented the most accurate, up to now, lattice determination 
of the second moment of the pion distribution amplitude using two 
flavors of dynamical (clover) fermions on lattices of different volumes 
and pion masses down to almost the physical value. So the chiral 
extrapolation \emph{per se}\ does not seem to be an issue.
Also the omission of strange quarks should not be of great importance.
However, the statistical fluctuations of the lattice matrix elements
of operators with derivatives are large for small pion masses and 
require averaging over a large number of configurations in order to 
obtain  phenomenologically relevant precision. We found that the 
signal can be somewhat improved by using the variational method with the
two interpolators corresponding to the pseudoscalar and axial-vector 
currents. 

The main difference of this work from the previous 
studies~\cite{Braun:2006dg,Arthur:2010xf} is the 
nonperturbative evaluation of the full $2\times2$ mixing matrix of 
the operators with two derivatives. In the framework of 
Ref.~\cite{Braun:2006dg} the nonperturbative mixing coefficient turns 
out to be of the same sign but up to one order of magnitude larger 
than the same coefficient computed perturbatively.
This observation underlines the necessity of nonperturbative
renormalization, at least at the presently reachable $\beta$-values.

Still, some uncertainty in the renormalization factors remains. It
is dominated by the uncertainty in the conversion factors connecting
the RI’-SMOM scheme to the $\MS$ scheme, which are calculated in
continuum perturbation theory and are known to two-loop 
accuracy~\cite{Gracey:2011fb, Gracey:2011zg}. A three-loop calculation is, 
therefore, needed in order to further reduce the renormalization 
uncertainty and would be extremely welcome.

In our work we have also emphasized the importance
of using the corrected relation 
Eq.~(\ref{a2-from-xi2a}) between $\langle \xi^2\rangle^\MS$ and 
$a_2^\MS$ for finite lattice spacing, instead of the continuum relation 
in Eq.~(\ref{a2-from-xi2}), due to discretization errors in derivatives 
that lead to a violation of the product rule. This effect is studied 
in detail.  

From our data we cannot exclude significant discretization effects in  
$\langle \xi^2\rangle^\MS$ and $a_2^\MS$, but a quantitative study 
requires simulations at smaller lattice spacings of the order of 
$a \sim 0.04 \, \mathrm {fm}$, which are presently not available to us. 
Such lattices will be generated in the future within the CLS 
effort~\cite{Bruno:2014jqa}. This will be a major step towards the 
calculation of the second moment of the pion DA with fully 
controllable accuracy. As a final remark, we note that the somewhat 
smaller value of $a_2^\MS$ obtained in this work seems to be 
favored by the phenomenological studies of form factors in the 
framework of light-cone sum rules, see, e.g., Refs.~\cite{Agaev:2010aq,Agaev:2012tm,Ball:2004ye,Duplancic:2008ix,Khodjamirian:2011ub}.  

\acknowledgments

This work has been supported in part by the Deutsche Forschungsgemeinschaft 
(SFB/TR 55) and the European Union under the Grant Agreement IRG 256594. 
The computations were performed on the QPACE systems of the 
SFB/TR 55, Regensburg's Athene HPC cluster, the SuperMUC system at the 
LRZ/Germany and J\"ulich's JUGENE using the Chroma software 
system~\cite{Edwards:2004sx} and the BQCD software~\cite{Nakamura:2010qh}  
including improved inverters~\cite{Nobile:2010zz,LuscherOpenQCD}.
We thank John Gracey for helpful discussions about renormalization
issues and the UKQCD collaboration for giving us permission to use 
some of their gauge field configurations.

\newpage

\begin{appendix}

\section{Bare and renormalized results by the ensemble}

The following Tables summarize the results obtained for each gauge field 
ensemble separately.

\begin{widetext}

\begin{table*}[ht!]
\renewcommand{\arraystretch}{1.2}
\caption{ \label{table:results_R-_R+_J5} Bare results for 
$R_{\rm av}^{\pm}$ using  $ J_5$ as interpolator.}
\begin{ruledtabular}
\begin{tabular}{cccclcclc}
$\beta$ & $\kappa_l$  & Size &
 Fit range  & $R_{\rm av}^{-}$ & $\chi^2/{\rm dof}$ & 
 Fit range  & $R_{\rm av}^{+}$ & $\chi^2/{\rm dof}$ \\
\hline
$5.20$&  $0.13596$  & $32^3 \times 64$ &    $3 - 14$& $0.1674 (36)$ &$0.67$ &$ 9 - 19$&$ 0.6013 (46) $&$6.67$ \\ 
$5.29$&  $0.13620$  & $24^3 \times 48$ &    $8 - 12$& $0.161  (25)$ &$0.47$ &$ 7 - 12$&$ 0.5792 (97) $&$4.44$ \\ 
$5.29$&  $0.13620$  & $32^3 \times 64$ &   $3 - 14$& $0.1668 (30)$ &$0.73$  &$11 - 20$&$ 0.6187 (52) $&$7.73$\\ 
$5.29$&  $0.13632$  & $32^3 \times 64$ &    $3 - 17$& $0.1705 (78)$ &$0.72$ &$ 9 - 15$&$ 0.602  (11) $&$7.76$ \\ 
$5.29$&  $0.13632$  & $40^3 \times 64$ &   $3 - 18$& $0.1756 (33)$ &$1.51$  &$10 - 25$&$ 0.6213 (36) $&$5.38$\\ 
$5.29$&  $0.13632$  & $64^3 \times 64$ &   $7 - 15$& $0.1694 (37)$ &$0.82$  &$15 - 26$&$ 0.6343 (22) $&$6.67$\\ 
$5.29$&  $0.13640$  & $64^3 \times 64$ &   $5 - 20$& $0.1627 (56)$ &$0.76$  &$17 - 25$&$ 0.6421 (61) $&$6.00$\\ 
$5.40$&  $0.13640$  & $32^3 \times 64$ &   $3 - 15$& $0.1679 (38)$ &$0.59$  &$14 - 25$&$ 0.654  (14) $&$3.83$\\ 
$5.40$&  $0.13647$  & $32^3 \times 64$ &   $3 - 15$& $0.1653 (35)$ &$1.04$  &$15 - 22$&$ 0.657  (21) $&$2.78$\\ 
$5.40$&  $0.13660$  & $48^3 \times 64$ &   $3 - 15$& $0.1681 (32)$ &$0.81$  &$15 - 25$&$ 0.6467 (57) $&$4.22$\\ 
\end{tabular}
\end{ruledtabular}
\renewcommand{\arraystretch}{1}
\end{table*}

\begin{table*}[ht!]
\renewcommand{\arraystretch}{1.2}
\caption{ \label{table:results_R-_R+_J45} Bare results for 
$R_{\rm av}^{\pm}$ using  $J_{45}$ as interpolator.}
\begin{ruledtabular}
\begin{tabular}{cccclcclc}
$\beta$ & $\kappa_l$  & Size &
 Fit range  & $R_{\rm av}^{-}$ & $\chi^2/{\rm dof}$ & 
 Fit range  & $R_{\rm av}^{+}$ & $\chi^2/{\rm dof}$ \\
\hline
$5.20$&  $0.13596$  & $32^3 \times 64$ &$10 - 16$ &$0.1859 (91) $ &$1.57 $  &$13 - 19$&$ 0.6354(72) $&$ 1.29$    \\ 
$5.29$&  $0.13620$  & $24^3 \times 48$ &$ 7 - 13$ &$0.1845 (83) $ &$1.08 $  &$10 - 15$&$ 0.680 (12) $&$ 0.92$    \\ 
$5.29$&  $0.13620$  & $32^3 \times 64$ &$10 - 15$ &$0.1963 (60) $ &$0.13 $  &$18 - 24$&$ 0.617 (10) $&$ 1.86$    \\ 
$5.29$&  $0.13632$  & $32^3 \times 64$ &$ 9 - 15$ &$0.155  (14) $ &$0.25 $  &$12 - 20$&$ 0.660 (13) $&$ 0.67$    \\ 
$5.29$&  $0.13632$  & $40^3 \times 64$ &$ 8 - 15$ &$0.1976 (49) $ &$0.83 $  &$17 - 24$&$ 0.6441(74) $&$ 1.19$    \\ 
$5.29$&  $0.13632$  & $64^3 \times 64$ &$10 - 25$ &$0.1839 (39) $ &$1.59 $  &$16 - 30$&$ 0.6394(23) $&$ 1.64$    \\ 
$5.29$&  $0.13640$  & $64^3 \times 64$ &$10 - 19$ &$0.2015 (97) $ &$0.77 $  &$20 - 30$&$ 0.6321(67) $&$ 0.37$    \\ 
$5.40$&  $0.13640$  & $32^3 \times 64$ &$ 7 - 15$ &$0.1931 (42) $ &$0.41 $  &$16 - 25$&$ 0.682 (13) $&$ 0.77$    \\ 
$5.40$&  $0.13647$  & $32^3 \times 64$ &$ 3 - 13$ &$0.1980 (17) $ &$0.59 $  &$17 - 22$&$ 0.682 (17) $&$ 0.22$    \\ 
$5.40$&  $0.13660$  & $48^3 \times 64$ &$14 - 20$ &$0.1823 (93) $ &$0.60 $  &$19 - 29$&$ 0.6640(73) $&$ 0.52$    \\ 
\end{tabular}
\end{ruledtabular}
\renewcommand{\arraystretch}{1}
\end{table*}

\begin{table*}[ht!]
\renewcommand{\arraystretch}{1.2}
\caption{ \label{table:results_Rpm} Bare results for $R_{\mathrm {av}}^{\pm}$ 
using  the variational method with the intrpolators $J_{45}$, $J_5$.}
\begin{ruledtabular}
\begin{tabular}{cccclcclc}
$\beta$ & $\kappa$  & Size &
 Fit range  & $R_{\mathrm {av}}^-$  &$\chi^2/{\mathrm {dof}}$ & 
 Fit range  & $R_{\mathrm {av}}^+$  &$\chi^2/{\mathrm {dof}}$ \\
\hline
$5.20$&  $0.13596$  & $32^3 \times 64$ &$3 - 16$ &$0.1813 (27)$ &$0.63 $ &$10 - 19$  &$0.6142 (46) $ & $0.52 $\\ 
$5.29$&  $0.13620$  & $24^3 \times 48$ &$3 - 13$ &$0.1660 (52)$ &$1.01 $ &$5  - 13$  &$0.6039 (54) $ & $0.38 $\\ 
$5.29$&  $0.13620$  & $32^3 \times 64$ &$4 - 16$ &$0.1775 (32)$ &$0.52 $ &$9  - 16$  &$0.6303 (35) $ & $0.41 $\\ 
$5.29$&  $0.13632$  & $32^3 \times 64$ &$6 - 16$ &$0.1710(120)$ &$0.63 $ &$5  - 16$  &$0.6289 (44) $ & $0.35 $\\ 
$5.29$&  $0.13632$  & $40^3 \times 64$ &$2 - 23$ &$0.1838 (24)$ &$1.52 $ &$14 - 24$  &$0.6226 (56) $ & $0.40 $\\ 
$5.29$&  $0.13632$  & $64^3 \times 64$ &$2 - 22$ &$0.1761 (21)$ &$0.85 $ &$8  - 25$  &$0.6353 (14) $ & $0.93 $\\ 
$5.29$&  $0.13640$  & $64^3 \times 64$ &$2 - 20$ &$0.1790 (39)$ &$0.78 $ &$10 - 20$  &$0.6350 (30) $ & $1.35 $\\ 
$5.40$&  $0.13640$  & $32^3 \times 64$ &$2 - 14$ &$0.1773 (27)$ &$0.55 $ &$13 - 20$  &$0.657  (11) $ & $0.45 $\\ 
$5.40$&  $0.13647$  & $32^3 \times 64$ &$2 - 16$ &$0.1742 (22)$ &$1.03 $ &$16 - 22$  &$0.662  (25) $ & $0.26 $\\ 
$5.40$&  $0.13660$  & $48^3 \times 64$ &$2 - 16$ &$0.1794 (24)$ &$0.80 $ &$15 - 25$  &$0.6534 (53) $ & $0.30 $\\ 
\end{tabular}
\end{ruledtabular}
\renewcommand{\arraystretch}{1}
\end{table*}

\end{widetext}

\begin{table}[ht!]
\renewcommand{\arraystretch}{1.2}
\caption{\label{table:I2} Results for 
$\langle 1^2 \rangle^{\MS}(\mu = 2 \, \mathrm {GeV})$  
using  the variational method with the interpolators $J_{45}$, $J_5$.
The first error corresponds to the statistical fluctuations, and the
second to the contribution from the uncertainty in the 
determination of the renormalization constants.}
\begin{ruledtabular}
\begin{tabular}{cccl}
$\beta$ & $\kappa$  & Size &
 $\langle 1^2 \rangle^{\MS}(\mu = 2 \, \mathrm {GeV} ) $  \\ 
\hline
$5.20$&  $0.13596$  & $32^3 \times 64$ & $0.9298 (70 ) (56)$   \\
$5.29$&  $0.13620$  & $24^3 \times 48$ & $0.9028 (81 ) (52)$   \\
$5.29$&  $0.13620$  & $32^3 \times 64$ & $0.9422 (53 ) (55)$   \\
$5.29$&  $0.13632$  & $32^3 \times 64$ & $0.9402 (66 ) (54)$   \\
$5.29$&  $0.13632$  & $40^3 \times 64$ & $0.9308 (84 ) (54)$   \\
$5.29$&  $0.13632$  & $64^3 \times 64$ & $0.9498 (20 ) (55)$   \\
$5.29$&  $0.13640$  & $64^3 \times 64$ & $0.9494 (44 ) (55)$   \\
$5.40$&  $0.13640$  & $32^3 \times 64$ & $0.9690 (159) (51)$   \\
$5.40$&  $0.13647$  & $32^3 \times 64$ & $0.9757 (371) (51)$   \\
$5.40$&  $0.13660$  & $48^3 \times 64$ & $0.9632 (79 ) (50)$   \\
\end{tabular}
\end{ruledtabular}
\renewcommand{\arraystretch}{1}
\end{table}

\begin{table}[ht!]
\renewcommand{\arraystretch}{1.2}
\caption{ \label{table:xi2} Results for 
$\langle \xi^2\rangle^{\MS} (\mu= 2 \, \mathrm {GeV})$
using  the variational method with the interpolators $J_{45}$, $J_5$.
The first error corresponds to the statistical fluctuations, and the
second to the contribution from the uncertainty in the 
determination of the renormalization constants.}
\begin{ruledtabular}
\begin{tabular}{cccl}
$\beta$ & $\kappa$  & Size &
 $\langle \xi^2\rangle^{\MS}(\mu = 2 \, \mathrm {GeV}) $ \\ 
\hline
$5.20$&  $0.13596$  & $32^3 \times 64$ & $0.2427(53 )  (28)$ \\
$5.29$&  $0.13620$  & $24^3 \times 48$ & $0.2147(103)  (42)$ \\
$5.29$&  $0.13620$  & $32^3 \times 64$ & $0.2325(63 )  (42)$ \\
$5.29$&  $0.13632$  & $32^3 \times 64$ & $0.2199(240)  (45)$ \\
$5.29$&  $0.13632$  & $40^3 \times 64$ & $0.2467(49 )  (38)$ \\
$5.29$&  $0.13632$  & $64^3 \times 64$ & $0.2289(42 )  (44)$ \\
$5.29$&  $0.13640$  & $64^3 \times 64$ & $0.2348(78 )  (42)$ \\
$5.40$&  $0.13640$  & $32^3 \times 64$ & $0.2284(58 )  (49)$ \\
$5.40$&  $0.13647$  & $32^3 \times 64$ & $0.2212(66 )  (51)$ \\
$5.40$&  $0.13660$  & $48^3 \times 64$ & $0.2334(50 )  (48)$ \\
\end{tabular}
\end{ruledtabular}
\renewcommand{\arraystretch}{1}
\end{table}

\begin{table}[ht!]
\renewcommand{\arraystretch}{1.2}
\caption{\label{table:a2} Results for $a_2^{\MS}(\mu = 2 \, \mathrm {GeV})$
using  the variational method with the interpolators $J_{45}$, $J_5$.
The first error corresponds to the statistical fluctuations, and the
second to the contribution from the uncertainty in the 
determination of the renormalization constants.}
\begin{ruledtabular}
\begin{tabular}{cccl}
$\beta$ & $\kappa$  & Size &
 $a_2^{\MS}(\mu = 2 \, \mathrm {GeV}) $ \\ 
\hline
$5.20$&  $0.13596$  & $32^3 \times 64$ & $ 0.1654  (161) (113)$ \\
$5.29$&  $0.13620$  & $24^3 \times 48$ & $ 0.0996  (304) (152)$ \\
$5.29$&  $0.13620$  & $32^3 \times 64$ & $ 0.1286  (188) (153)$ \\
$5.29$&  $0.13632$  & $32^3 \times 64$ & $ 0.0930  (700) (161)$ \\
$5.29$&  $0.13632$  & $40^3 \times 64$ & $ 0.1767  (160) (141)$ \\
$5.29$&  $0.13632$  & $64^3 \times 64$ & $ 0.1136  (124) (158)$ \\
$5.29$&  $0.13640$  & $64^3 \times 64$ & $ 0.1310  (230) (154)$ \\
$5.40$&  $0.13640$  & $32^3 \times 64$ & $ 0.1010  (214) (169)$ \\
$5.40$&  $0.13647$  & $32^3 \times 64$ & $ 0.0760  (380) (176)$ \\
$5.40$&  $0.13660$  & $48^3 \times 64$ & $ 0.1188  (159) (164)$ \\
\end{tabular}
\end{ruledtabular}
\renewcommand{\arraystretch}{1}
\end{table}

\end{appendix}

\end{document}